\documentclass[useAMS,usenatbib]{mn2e}
\usepackage{graphicx,epsfig,amssymb,amsmath,layout,float,verbatim,rotating,flafter,enumerate,calc}

\title[Chaotic motion and spiral structure]
{Chaotic motion and spiral structure in self-consistent models of
rotating galaxies}
\author[N. Voglis, I. Stavropoulos and C. Kalapotharakos]{N. Voglis$^{1}$\footnotemark[1], I. Stavropoulos$^{1,2}\footnotemark[1]$ and C. Kalapotharakos$^{1}$
\thanks{e-mail:
nvogl@cc.uoa.gr (NV); istavrop@phys.uoa.gr (IS);
ckalapot@phys.uoa.gr (CK}
\\ $^{1}$Academy of Athens, Research Center for Astronomy, 4 Soranou Efesiou Str.,
GR-11527
\\ $^{2}$University of Athens, Department of Physics, Section of Astrophysics}

\date{Accepted ..........Received .............;in original form ..........}

\pagerange{\pageref{firstpage}--\pageref{lastpage}}

\pubyear{2006}

\label{firstpage}

\begin{document}

\maketitle

\begin{abstract}
Dissipationless N-body models of rotating galaxies, iso-energetic
to a non-rotating model, are examined as regards the mass in
regular and in chaotic motion. Iso-energetic means that they have
the same mass and the same binding energy and they are near the
same scalar virial equilibrium, but their total amount of angular
momentum is different. The values of their spin parameters
$\lambda$ are near the value $\lambda=0.22$ of our Galaxy.

We distinguish particles moving in regular and in chaotic orbits
and we show that the spatial distribution of these two sets of
particles is much different. The rotating models are characterized
by larger fractions of mass in chaotic motion (up to the level of
$\thickapprox 65\%$) compared with the fraction of mass in chaotic
motion in the non-rotating iso-energetic model (which is on the
level of $\thickapprox 32\%$). Furthermore, the Lyapunov numbers
of the chaotic orbits in the rotating models become by about one
order of magnitude larger than in the non-rotating model. This
impressive enhancement of chaos is produced, partly by the more
complicated distribution of mass, induced by the rotation, but
mainly by the resonant effects near corotation. Chaotic orbits are
concentrated preferably in values of the Jacobi integral around
the value of the effective potential at the corotation radius.

We find that density waves form a central rotating bar embedded in
a thin and a thick disc with exponential mean radial profile of
the surface density. A surprising new result is that long living
spiral arms are excited on the disc, composed almost completely by
chaotic orbits.

The bar excites an $m=2$ mode of spiral waves on the surface
density distribution of the disc, emanating from the corotation
radius. The bar goes temporarily out of phase with respect to an
excited spiral wave, but it comes in phase again in less that a
period of rotation. As a consequence, spiral arms show an
intermittent behavior. They are deformed, fade, or disappear
temporarily, but they grow again re-forming a well developed
spiral pattern. Spiral arms are discernible up to 20 or 30
rotations of the bar (lasting for about a Hubble time). The
relative power of the spiral $m=2$ mode with respect to all other
fluctuations on the surface density is initially about $50\%$, but
it is reduced by a factor of about 2 or 3 at the end of the Hubble
time.

\end{abstract}


\section{Introduction}
In previous papers some cases of non-rotating N-body models of
elliptical galaxies have been investigated as regards their mass
components in ordered and in chaotic motion (Voglis,
Kalapotharakos \& Stavropoulos 2002; Kalapotharakos, Voglis \&
Contopoulos 2004; Kalapotharakos \& Voglis 2005). In these studies
the fraction of mass in chaotic motion and its consequences were
found to be important.

In particular, in Voglis et al. (2002) two non-rotating triaxial
equilibrium models, called C and Q, resulting respectively  from
clumpy (C) and quiet (Q) cosmological initial conditions after
dissipationless collapse and relaxation, are examined. Both models
have smooth centers, i.e. the density near the center flattens
inside a radius of about 10\% of the half mass radius. The C model
is more close to spherical than the Q model having a maximum
ellipticity E4. The mass in chaotic motion in C is found to be
about $24\%$ of the total mass. The Lyapunov numbers of the orbits
are found in general small, so that only $2\%$ of the total mass
is able to develop chaotic diffusion in a Hubble time. The Q model
has a high maximum ellipticity E7. The fraction of mass in chaotic
motion in Q is found to be about $32\%$ of the total mass and the
values of the Lyapunov numbers are found higher than in C, so that
a fraction of mass about $8\%$ is able to develop chaotic
diffusion in a Hubble time.

We have also found that the spatial distribution of the mass in
chaotic motion is different than the spatial distribution of the
mass in ordered motion, so that this difference can form an
observable hump in the profile of the surface density.

The fact that chaos in models of non-rotating galaxies with smooth
centers is not negligible was also found in Muzzio, Carpintero \&
Wachlin (2005). These authors found a fraction of mass in chaotic
motion of about $53\%$ in their non-rotating smooth center model,
i.e. even larger than the fraction found in our models. One reason
that explains this difference is the fact that their model is more
cuspy, i.e. the density near the center flattens in a smaller
radius than in our models. Another reason is that they have used a
lower threshold of the Lyapunov numbers for distinguishing between
regular and chaotic orbits. The two results are in agreement, in
the sense that the fraction of mass in chaotic motion cannot be
negligible (it is of the order of some tens percent) even in
models with smooth centers that are expected to be mostly in favor
of ordered motion. Another agreement with our results is that the
mass in chaotic motion has a different spatial distribution than
the mass in regular motion. However, the Lyapunov times of the
vast majority of chaotic orbits are large compared with the mean
dynamical times of galaxies (e.g. more than 10 dynamical times).

Models of non-rotating galaxies are the simplest possible models
of galaxies, especially those with smooth centers. Rotation of
galaxies introduces new parameters and nonlinear interactions that
lead to more complicated dynamics. In rotating models density
waves, able to form rotating bars or spiral arms, can grow.

The mechanism that is responsible for the formation and the
stability of spiral arms in galaxies is a complicated problem that
has not been very well understood yet. All researchers agree that
the main driving force leading to this effect is gravity.
Dissipative effects can also play a non-negligible role in the
stability of the spiral arms, but gravity remains the main driving
mechanism.

The frequent appearance of spiral arms in galaxies is in favor of
the assumption that this is probably a quasi-stationary effect,
i.e. spiral arms are long living features in galaxies. This
assumption is known as the Lin-Shu Hypothesis (Lin \& Shu
1964,1966; Binney \& Tremaine 1987).

The dynamical evolution of galaxies can be studied by combining
theoretical analysis with the results obtained by N-body
simulations.

N-body simulations show that spiral arms appear as a transient
effect, in the sense that they grow as density waves, but they are
temporarily dissolved or fade and recreated again. This process
may be repeated several times until they finally disappear.

A necessary ingredient for spiral structure to appear is the
amount of angular momentum a galaxy possesses, as well as the way
this angular momentum is distributed along the mass of the galaxy.

It is well known that resonances between the frequencies of
stellar orbits and the angular velocity of the pattern $\Omega_p$
play a significant role on the stability and the dynamical
evolution of rotating discs. In particular the Inner Lindblad
Resonance (ILR), Corotation (CR) and the Outer Lindblad Resonance
(OLR) are the most important resonances. Lynden-Bell \& Kalnajs
(1972) have shown that in a rotating disc angular momentum is
transferred outwards. This transference is due to the interaction
of a bar-like perturbation of the gravitational field with stars
moving in resonant orbits particularly near ILR and CR. Tremaine
\& Weinberg (1984) and Weinberg (1985), extended this analysis
showing that angular momentum transference can be interpreted as
an effect caused by a particular type of dynamical friction.

N-body simulations indicate that angular momentum transference
outwards is always  present in rotating galactic models. This
effect takes place not only between the bar and the disc, but it
is extended to the halo particles (e.g. Debattista \& Sellwood
1998, 2000; Athanassoula 2002, 2003).

In particular, it has been found that the material in the bar and
in the inner disc emits angular momentum to resonant material in
the outer disc, or in the halo and in the bulge. The evolution of
bars in isolated disc galaxies is driven by the redistribution of
angular momentum. Transference of angular momentum outwards in
rotating galaxies produces secular dynamical evolution in which
the bar grows and slows down.

Several questions arise regarding the role of regular motion and
the role of chaotic motion in rotating galaxies and their
evolution, examples are as follows.

 \noindent
i) What is the effect of rotation on the degree of chaos in such
systems, in comparison to the degree of chaos in non-rotating
systems?

\noindent
 ii) What is the typical level of the Lyapunov Characteristic
 Numbers  and what is the amount of mass that can develop efficient
 chaotic diffusion in a Hubble time?

\noindent
 iii) What is the role of chaos on the secular
 evolution and how secular evolution affects chaos?

\noindent
 iv) How the two components of mass in chaotic or in regular motion are
 distributed in space?

\noindent
 v)How the pattern is affected by chaos? Can such systems develop spiral arms? For how long?

In order to investigate these questions, we use one of the
non-rotating N-body models mentioned above, namely the Q-model,
and we construct from it a series of iso-energetic rotating
models. Iso-energetic means that they have the same total mass
$M$, the same total binding energy $E$ and they are close to the
same scalar virial equilibrium, but they carry different total
amounts of angular momentum. In this way, a direct comparison can
be made as regards the effect of their angular momentum on the
behavior of the models. The total amount of angular momentum in
these models is large, so that their values of the spin parameter
$\lambda$ are comparable to the value of $\lambda$ of our Galaxy.

In this investigation we use the same code and the same number of
particles as in Voglis et al. (2002), i.e. an improved version of
the conservative technique code by Allen Palmer \& Papaloizou
(1990) with $N \thickapprox 1.3 \times 10^5$ particles.

We use as a scaling unit of length the half mass radius $R_h$ in
every model. The half mass crossing time
$T_{hmct}=(2R_h^3/GM)^{1/2}$ is defined as the unit of time. An
important time scale is the radial period $T_{hmr}$ of an orbit
with binding energy $E_{hmr}$ equal to the mean value of the
potential at the half mass radius. If we assume that in a typical
galaxy a star with binding energy $E_{hmr}$ describes about 50
circular periods $T_{cp}=2 \pi T_{hmct}$ in a Hubble time, then it
describes about 100 radial periods in this time (since the 2:1
resonance at this energy is typical). Therefore, a Hubble time is
$t_{Hub} \approx 50 T_{cp} \approx 100 T_{hmr} \approx 300
T_{hmct}$.

In Section 2 the method of derivation of the initial conditions of
our models is described. In Section 3 various features of the
models along their self-consistent run are examined, regarding
their morphology, the radial profiles of the surface density, the
rotation curves, the evolution of the pattern speed and the
corotation radius, the longevity of the spiral structure and the
secular evolution of the systems. In Section 4 the components of
mass in regular and in chaotic motion for every model are
estimated by the method introduced in Voglis et al. (2002). In
Section 5 the spatial distribution of these two components is
demonstrated. Our conclusions are summarized and a discussion is
given in Section 6.

\section{Initial conditions for the rotating models}

The Q model is a non-rotating triaxial equilibrium configuration.
The three components of the total angular momentum, $J_x$, $J_y$,
$J_z$ along the principal axes of the system are very close to
zero. The ratio of the direct to the retrograde orbits with
respect to any axis is very close to $1:1$.

As described below, rotation is inserted by re-orienting the
velocities of particles, so that all the orbits rotate initially
around the same axis and along the same direction. (Of course, one
can create models by only a partial initial re-orientation of the
velocities). This is a kind of ``conspiracy" between orbits,
introduced artificially, that increases the total amount of
angular momentum in the system, but preserves the kinetic energy.

A measure of the angular momentum $J$, for a gravitational system
of total mass $M$ and total binding energy $E$, can be given in
terms of the spin parameter $\lambda$ introduced by Peebles
(1969). This parameter is defined by the equation
\begin{equation}
\lambda=\frac{J|E|^{1/2}}{GM^{5/2}}
\end{equation}
Notice that the value of $\lambda$ of the disc of our Galaxy is
$\lambda_{Galaxy}=0.22$, as  has been evaluated by Efstathiou \&
Jones (1979)  from the value of the angular momentum of our Galaxy
given by Innanen (1966).

The conspiracy of the directions of rotation of the orbits
mentioned above is important for the formation of fast rotating
patterns and gives to the spin parameter $\lambda$ values
comparable to the value of our Galaxy.

It should be stressed here that in a system with random directions
of rotation of the orbits the total angular momentum of the system
is very small, or zero. Even if the orbits rotate around the same
axis, but those in direct rotation are comparable and mixed with
the orbits in retrograde rotation, the total angular momentum of
the system is again small, or zero. Pattern rotation in such
models is negligible, or zero.

If galaxies are formed from small initial density perturbations in
a hierarchical clustering cosmological scenario and they have
acquired their angular momentum from tidal interactions with other
density perturbations of their environment in the early Universe,
a conspiracy between the directions of rotation of the orbits
cannot be obtained in a dissipationless scenario either in
monolithic collapses, or in mergers between sub-clumps formed
hierarchically. Tidal interactions between density perturbations
can only give structures with a mean value of the spin parameter
of about $\lambda=0.05$ (e.g. Efstathiou \& Jones 1979; Barnes \&
Efstathiou 1987; Zurek, Quinn \& Salmon 1988; Voglis \& Hiotelis
1989; Voglis, Hiotelis \& H\"{o}flich 1991). The main reason for
such a small value of $\lambda$ is the mixing between direct and
retrograde orbits.

Various estimations show that in a galaxy formation process a
value of $\lambda$ near 0.2 can be achieved, if the collapse
factors are remarkably larger than the value of 2, that results
from the virial theorem, when it is applied to a dissipationless
collapse process. Such a large value of the collapse factor can be
obtained by dissipative processes. Furthermore, dissipative
processes absorb the retrograde motion (by collisions between gas
clouds) and lead to systems where direct motion dominates.

Therefore, according to this scenario, only the dissipative mass,
i.e. the baryonic component, of a galaxy can be characterized by
large values of $\lambda$. This component is mostly related to the
disc of galaxies. In principle, a dark matter halo should be
characterized by small values of $\lambda$, compatible with those
produced by cosmological tidal fields, as mentioned above.

In the present paper  we compare iso-energetic fast rotating
models of various angular momenta with a non-rotating model, in
order to reveal the consequence of rotation as regards the level
of chaos. For this reason, in this paper, we use a simple model
where no halo component with low values of $\lambda$ is
considered. As we will see in the next section in our rotating
systems a bar and a thin disc embedded in a thick disc are formed.
An extension of this work combining a discy component of large
$\lambda$ with a halo component of small $\lambda$ will be
presented elsewhere.

In order to create initial conditions for an iso-energetic
rotating model based on the Q model, we apply the following
velocity re-orientation process:

At a given snapshot, at $t_Q=100$, of the self-consistent
evolution of the Q model, the component $v_{yz}$ (parallel to the
plane of intermediate-longest axes) of the velocity of every
particle is re-oriented, so that it becomes perpendicular to the
current radius $r_{yz}$ of the particle on this plane, pointing
along the same (clockwise) direction.

This velocity re-orientation process does not affects the total
kinetic energy $T$  and respects the scalar viral equation of the
system
\begin{equation}
T = T_{rot,yz}+ T_{rad,yz}+ T_x = -E
\end{equation}
where $T_{rot,yz}$ and $T_{rad,yz}$ are the rotational and the
radial kinetic energies parallel to the $y-z$ plane. $T_x$ is the
kinetic energy along the (shortest) $x$ axis.

The initial conditions created in this way form a rotating new
model called QR1. At the time $t_{QR1}=0$, when the running of
this model starts,  the radial components of the velocities
parallel to the y-z plane are zero. All the kinetic energy
parallel to this plane is rotational kinetic energy around the
shortest axis (x), of the same direction. As a consequence, the
total angular momentum of the system is equal to a non-zero
component $J_{x}$. During the (self-consistent) evolution of this
model, radial motion grows rapidly, but the total angular momentum
$J_x$ of the system remains constant.

In order to create initial conditions for the next model, QR2, we
choose a snapshot of the self-consistent run of the QR1 model at
$t_{QR1} \approx 20$. This period of time is long enough for some
very fast transient effects of the QR1 model, due to the
redistribution of the types of orbits, to decay. During this
period considerable amount of radial motion has been grown. The
initial conditions of the new model QR2 are created by applying
again the velocity re-orientation process at this snapshot of the
QR1 model. The time is reset to $t_{QR2}=0$, when the running of
QR2 starts. The total angular momentum $J_x$ becomes larger in QR2
than in QR1.

Initial conditions of the model QR3 are created in a similar way
by using a snapshot of the self-consistent run of the QR2 model at
$t_{QR2} \approx 20$ on which the same velocity re-orientation
process is applied. The time of QR3 is reset to $t_{QR3}=0$, when
its running starts. Angular momentum $J_x$ is larger in QR3 than
in QR2.

The initial conditions of the model QR4, with angular momentum
$J_x$ larger  that in QR3, are created from the model QR3 in a
similar way. Again the time is reset to $t_{QR4}=0$, at the start
of its running.

The values of $\lambda$ of the above systems corresponding to the
mass inside a cylinder with radius $r=3$ on the equatorial plane,
for the four models respectively, are $\lambda \thickapprox
0.21,~~0.25,~~0.28,~~0.30$. Thus, we have a sequence of five
iso-energetic models, Q, QR1, QR2, QR3, QR4, with different
amounts of their angular momentum $J_x$.

Before the final runs a number of testing runs were performed with
various time steps $\Delta t$ at which the self-consistent
potential is re-evaluated. In all the runs a pronounced rotating
bar is formed by density waves, as it is described in the next
section. We found that the rotation of the bar is sensitive in the
choice of $\Delta t$. Namely, if $\Delta t$ is not small enough
the bar suffers from a systematic numerical deceleration. In
particular as the orbits are integrated during $\Delta t$ in a
constant potential they form a new bar at a small phase $\Delta
\Phi$ ahead with respect to the bar of the potential. This lack of
self-consistency for the time $\Delta t$ introduces a numerical
retarding torque that causes an unrealistic slowing down of the
bar. The cumulative effect of this torque for a Hubble time can be
serious, unless $\Delta t$ is small enough. We found that $\Delta
t =5 \times 10^{-4} T_{hmct}$ is a good choice in order to
minimize the hysteresis between the bar of the potential and the
bar of the real density of particles during $\Delta t$, so that
the cumulative effect in a Hubble time remains small.

Every model is run self-consistently for a Hubble time, i.e. a
time $t=300$ in our time units. During this run the coefficients
of the self-consistent potential are re-evaluated at every $\Delta
t =5 \times 10^{-4} T_{hmct}$. The cumulative variation of the
total energy during the whole run is less than ${\Delta E \over
E}=0.4\%$. In the rotating models a number of particles about
$0.5\%$ acquire positive energies and escape the system. The
center of mass, being initially at $r=0$ is translated by a
distance less than 0.03.

\section{Morphological features and secular evolution}
\subsection{Morphology}
During the self-consistent evolution, in the rotating models,
density waves grow to form a strong bar surrounded by an inner
disc, an outer disc and a thick disc. (The density of the disc
decreases with increasing $\mid x \mid$, so that a thin disc can
be defined near the equatorial plane y-z). The bar is triaxial and
rotates around its shortest axis x.

In the first row of Fig. 1 the edge-on projections (y-x plane) of
the four models (QR1, QR2, QR3, QR4) are shown, respectively, at
their snapshots at time $t_1 \approx 20$ of their run time. (More
precisely, $t_1$ is $20,~~21,~~18,~~18$ for the four models,
respectively. Hereafter, these times are referred simply as
$t_1\approx 20$).

In the second row of Fig. 1 the face-on projections (on the
equatorial plane, y-z) of these models are shown at the same
snapshots as in the first row. In these figures we can see that
the size of the bar decreases with increasing angular momentum
along the sequence of the models. We can also see that density
waves form two symmetric spiral arms emanating from the ends of
the bar, that become more prominent along the sequence QR1, QR2,
QR3, QR4.

The third and the forth row in Fig. 1 are similar to the first two
rows, respectively, but at a time of about $100 T_{hmct}$ later.
During this long interval of time ($t=0\rightarrow t\approx 120$)
the bar has described a total number of about 9, 11, 13, 15
rotations, respectively in the four models. We see that in QR4
spiral arms are still well discernible at this time. In QR2 and
QR3 they are fainter. In QR1 they cannot be seen in this
projection. Furthermore, we see that the distribution of mass in
QR2, QR3, QR4 has been expanded along the equatorial plane by a
factor that is more prominent in the QR4 model.

\subsection{Radial profiles of the surface density}
In the five frames of the left column of Fig. 2 the radial
profiles of the mean surface density  $\overline{\sigma}_{yz}(r)$
at a radius $r$ on the equatorial plane of all the models (Q, QR1,
QR2, QR3, QR4) are shown as functions of the radius $r$, in
$r$-$\log_{10}{\overline{\sigma}_{yz}}(r)$ scale.

As the rotation increases along the sequence of the models, the
radial surface density profiles at small radii ($r \lesssim 1$)
become steeper. In the rotating models the slope in the outer
parts remains remarkably constant along the radius, indicating an
exponential radial profile of the surface density.

The time variations of these profiles during a Hubble time are
small. For example, in the frames for the models QR1,QR2,QR3,QR4,
in Fig. 2 the surface density radial profiles are drawn at $t_1
\thickapprox 20$ (solid line) and at $t_2=t_{Hub}=300$ (dashed
line). There is a tendency of these profiles to become flatter in
time. This tendency increases along the sequence of the models and
it is due to a slow expansion of the mass at large radii combined
with a slow contraction at smaller radii.

The exponential radial profiles of the surface density in the
rotating models can also be seen in the five frames of the right
column of Fig. 2 where the corresponding slopes
\begin{equation}
k=-\frac{d \log_{10} \overline{\sigma}_{yz}(r)}{dr}
\end{equation}
are shown as functions of $r$ at an intermediate snapshot of
$t=150$. We see that the value of the slope $k$ in the
non-rotating Q model decreases almost constantly up to large
radii. However, in the rotating models QR1, QR2, QR3, QR4 the
values of $k$ show a maximum at a small radius ($r < 1$), but
beyond this radius, $k$ flattens and remains remarkably constant
up to large radii, expressing the exponential character of the
profile. Such radial profiles of the surface density are similar
to the observed profiles of the discs (see, for example, Grosbol
\& Patsis 1998, figs.1a-c therein).

\subsection{Rotation curves and angular momentum transference}
It is interesting to examine the time evolution of the rotational
velocity curves and the distribution of the angular momentum for
each model.

In order to find the rotation curve at a given snapshot of such a
system we define 50 successive cylindrical co-axial rings around
the rotation axis (x), so that the mass inside every ring is the
same for all the rings. The rotational velocity $v_i$ at the mean
radius $r_i$ of the ring $i$ is evaluated from the equation
\begin{equation}
v_i={J_i \over r_i}
\end{equation}
where
\begin{equation}
J_i= \frac{1}{N_i}\sum_{j=1}^{N_i}(y_j \dot{z_j}-\dot{y_j}z_j)
\end{equation}
is the mean angular momentum per particle along the x axis of the
$N_i$ particles in the ring $i$.

In the left column of Fig. 3 the rotational velocity curves of the
four models are shown. In these figures the velocity $v$ is
measured in units of $R_h/T_{hmct}$. The conversion factor in real
units is
\begin{equation}
v_{real units}=v*21.5 {\text{km} \over \text{sec Kpc}} R_h
\end{equation}
The value of $R_h$ in $\text{Kpc}$ depends on the cosmological
model, on the total mass of the system and on the value of the
redshift $z_v$ at the epoch of virialization of the galaxy. It is
given by
\begin{equation}
R_h \approx 8 \text{Kpc}(\Omega h^2)^{-1/3}{9 \over 1+z_v} ({M
\over M_{12}})^{1/3},
\end{equation}
where $\Omega$ is the total density parameter of the Universe, $h$
is the Hubble constant in units of $100 {\text{Km} \over \text{sec
Mpc}}$, $M$ is the mass of the galaxy and $M_{12}=10^{12}
M_{\odot}$. For $\Omega=1$, $h=0.72$, $M=5*10^{11} M_{\odot}$ and
a redshift of formation $z_v=5$ we get
\begin{equation}
R_h \approx 10.6 \text{Kpc}
\end{equation}
which is a typical value for $R_h$.

The solid lines in Fig. 3 give the rotational velocity curves
$v_i(r_i)$ evaluated from (4) at different times, namely
$t=0,~20,~300$, as indicated by the numbers next to the curves.

For comparison, dashed lines in these figures give the
``dynamical" rotational velocity $v(r)=\sqrt{r\dfrac{dV(r)}{dr}}$
curves as evaluated at $t_1 \approx 20$ and at $t_2 \approx 300$
from the centrifugal equilibrium under the forces
$\dfrac{dV(r)}{dr}$ due to the monopole terms of the potential. At
the end of a Hubble time the values of $v(r)$ become slightly
larger near the center, while they become slightly smaller at
large radii, indicating a slow redistribution of the mass during
this time.

However, this variability of $v(r)$ is small compared to the
evolution of the rotational velocity curves $v_i(r_i)$ in all the
models. In the inner parts (e.g. for $r \lesssim 1$) the
rotational velocity $v_i(r_i)$ becomes gradually smaller. This
occurs at a decreasing rate and it is a result due to the
transference of angular momentum outwards.

The transference of angular momentum outwards can be seen in the
right column of Fig. 3, where the cumulative angular momentum is
plotted as a function of the cumulative mass from the center at
the same snapshots as the rotation curves in the left column of
Fig. 3. We see that at later times more mass is required in order
to collect the same amount of angular momentum. It is remarkable
that the transference of angular momentum outwards depends also on
the model. For example, in QR1 the distribution of angular
momentum at $t_1=20$ is close to the distribution at $t_2=300$,
while in QR2, QR3 and QR4 the distribution of angular momentum at
$t_1=20$ is close to the distribution at $t=0$.

\subsection{Evolution of the pattern speeds and the corotation radii}
A direct consequence of the angular momentum transference is that
the bar grows and its rotation slows down. We examine the
deceleration of the bar, i.e. the evolution of its pattern speed
and the corresponding corotation radius.

We find first, numerically, the function of time $\Phi(t)$, that
gives the angle $\Phi$ of the orientation of the longest axis of
the inertia ellipsoid of the mass inside a radius $r$ (containing
the bar), with respect to the initial orientation of this axis.
The pattern speed of the bar as a function of time is evaluated
from the slope of $\Phi(t)$ at time $t$, i.e.
$\Omega_p(t)=\dfrac{\Delta \Phi(t)}{\Delta t}$.

The corotation radius at a particular snapshot can be estimated by
locating the unstable Lagrangian point $L_1$ from the autonomous
Hamiltonian (Jacobi integral)
\begin{equation}
\label{3Dham}
H=\frac{\dot{r}^{2}}{2}+\frac{P_{\phi}^{2}}{2r^{2}}-\Omega_p
P_{\phi}+ \frac{\dot{x}^{2}}{2}+ V(\sqrt{r^2+x^2}) + V_1(r,\phi,x)
\end{equation}
where $(r, \phi)$ are the polar coordinates on the equatorial
plane (yz) and $ \Omega_p $ is the value of the pattern speed of
the bar at this snapshot. In this expression the term
$V(\sqrt{r^2+x^2})$ is the monopole component of the
self-consistent potential at this snapshot.

In Fig. 4 the time evolution of $\Omega_p(t)$ (Fig. 4a) and of the
corresponding corotation radius $r_c$ (the radius of the unstable
Lagrangian point $L_1$, Fig. 4b) are plotted for every model.

In QR1 $\Omega_p(t)$ does not vary considerably preserving a value
about $0.4$ (in units of $rads/T_{hmct}$). The corotation radius
also remains roughly constant at about 2.3.

In QR2, QR3, QR4 $\Omega_p(t)$ starts from larger values about
0.75, 0.90, and 1.0, respectively, and decreases initially faster,
but more slowly later on. The corotation radii start from the
values about 1.3, 1.2, 1.0 and increase by about $50\%$ at the end
of the Hubble time. Gadotti \& de Souza (2005, 2006) investigated
the kinematics, the lengths, and the colors and ages in a good
number of face-on bars in real galaxies. They classified the bars
in these galaxies in old and young bars. One of their main results
is that bars can survive for a Hubble time. Furthermore, the mean
linear size of the young bars is $5.4 \pm 1.6 \text{Kpc}$, while
the mean linear size of the old bars is $7.5 \pm 1.2 \text{Kpc}$,
i.e. old bars are by about $40\%$ longer than the young bars. Thus
our results are in agreement with observations, as regards the
time evolution of the linear sizes of the bars.

\subsection{On the longevity of spiral arms}
The surface density $\sigma_{yz}(r,\phi,t)$ of the mass projected
on the equatorial plane is a function of the position $(r,\phi)$
on this plane and of the time $t$. An important question regards
the evolution and the longevity of the spiral pattern. In order to
investigate this question we define the relative surface density
fluctuation of $\sigma_{yz}(r,\phi,t)$ with respect to the mean
surface density $\overline{\sigma}_{yz}(r,t)$ at the radius $r$,
i.e.
\begin{equation}
\delta(r,\phi,t)=\frac{\sigma_{yz}(r,\phi,t)-\overline{\sigma}_{yz}\,(r,t)}{\overline{\sigma}_{yz}\,(r,t)}
\end{equation}
and we analyzed $\delta(r,\phi,t)$ in Fourier component as
\begin{equation}
\delta(r,\phi,t)=\sum_m A_m(r,t)\cos{m[\phi-\phi_m(r,t)]}
\end{equation}
The appearance of two spiral arms requires a serious relative
contribution of the $m=2$ mode (with respect to the total power of
$\delta(r,\phi,t))$ and a monotonic (on the average, increasing or
decreasing) function of $\phi_2(r,t)$ with $r$, at least beyond a
certain (not very large) radius.

The relative contribution of the $m=2$ mode is measured by
\begin{equation}
a_2^2(r,t)=\frac{A_2^2}{C^2}
\end{equation}
where
\begin{equation}
C^2=\sum_m A_m^2
\end{equation}
We define 80 successive cylindrical co-axial rings around the
rotation axis (x), extended up to $r=5$, with equal width $\Delta
r$. We calculate the values of $a_2^2(r,t)$  and the corresponding
azimuthal angle $\phi_2(r,t)$ at any given time $t$.

A series of 6 snapshots are displayed in Fig. 5 for every model
QR1, QR2, QR3, QR4 at various times along the evolution for a
Hubble time. For every model two columns are displayed. The left
column shows the distribution of values of
\begin{equation}
\delta_2(r,\phi,t)=A_2(r,t)\cos{2[\phi-\phi_2(r,t)]}
\end{equation}
on the equatorial plane $(r, \phi)$. In these figures black means
positive and white means negative values of $\delta_2(r,\phi,t)$.
We see that the $m=2$ mode forms a pattern composed of two parts.

The first part is a `bow tie' pattern around the center, where the
phase $\phi_2$ is independent of $r$ and corresponds to the
central part of the bar, composed mainly of regular orbits (see
Section 4). This `bow tie' pattern ends at a radius shorter than
the corotation radius, close to the Inner Lindblad resonance
(ILR).

An example of how the three resonances (ILR, CR, OLR) are located
in the distribution of $\delta_2(r,\phi,t)$ is given in Fig. 6.
The snapshot of QR4 at $t=55$ (second row, seventh column of Fig.
5) is shown in magnification in Fig. 6, with three cycles drawn at
the radii of the above resonances.

The second part in the distribution of $\delta_2(r,\phi,t)$ is a
two-armed spiral pattern emanating from the ends of the bar near
the corotation radius. This structure appears in all the models
and, despite its intermittent behavior described below, it is
recreated up to the end of the Hubble time for all the models.
Notice that, during the Hubble time the bar has described about
19, 25, 29 and 33 rotations in the four models QR1, QR2, QR3, QR4,
respectively.

The right column for each model in Fig. 5 gives the distribution
of $\delta_{3imp}(r,\phi,t)$ on the equatorial plane,  as it comes
out from the superposition of the three most important modes (i.e.
the modes with the three highest amplitudes) of the spectrum at
the same snapshots as the left column. Most frequently and for the
majority of the rings outside the corotation radius, three most
important modes are those with $m = 2, 4, 6 $. Sporadically at
various rings and occasionally in time the modes with $m = 3, 5,
8, 10, 12$ appear also in $\delta_{3imp}(r,\phi,t)$.

A careful examination of the figures in the right column of each
model in Fig. 5 shows that, although the distribution of the $m=2$
mode is to some extent obscured by the other modes, it is still
discernible even at the latest snapshots (see for example the last
row in Fig. 5, where the snapshots at $t=288, ~~280,~~ 294,~~ 291$
for the four models are displayed).

As regards the dependence of the azimuthal phase $\phi_2(r,t)$ on
the radius $r$ along the spiral arms this is very close to linear,
on the average. This can be seen if we plot the distributions of
$\delta_2(r,\phi,t)$ and $\delta_{3imp}(r,\phi,t)$ on a
rectangular frame $(r,\phi)$ as shown in Fig. 7 for the snapshots
of the four models shown in the second row of Fig. 5, for example.
The spiral arms are displayed in this figure by dark lanes along
almost straight oblique lines beyond the corotation radius up to a
radius $r=5$, indicating that $\phi_2(r,t)$ is approximately
linear in $r$ along the spiral arms.

Fig. 8 is the same as Fig. 7, but for the latest snapshots of the
four models shown in the last row of Fig. 5. In these late
snapshots of evolution the linear dependence of $\phi_2(r,t)$ on
$r$ is shorter, extended from the corotation radii to a radius of
about $r=3$ or $3.5$.

In order to have some idea of the absolute size of the $m=2$ mode
at the early and the late stages of evolution, we examine the
amplitude $A_2(r,t)$ of $\delta_2(r,\phi,t)$ in (14). As an
example, in Fig. 9 we give $A_2(r,t)$ as a function of $r$ at the
two snapshots of QR4 shown in the second row ($t=55$) and the last
row ($t=291$) of Fig. 5. In Fig. 9a (snapshot at $t=55$) the
amplitude $A_2(r,t)$ has a maximum value about 1.1 inside the bar
($r<1$) and falls to a level of about 0.4 in the region of spiral
arms. In Fig. 9b ($t=291$) the maximum value of $A_2(r,t)$ in the
bar is only slightly reduced. In the region of spiral arms,
however, it falls to the level of about 0.1 or 0.2.

Ohta, Hamabe \& Wakamatsu (1990) estimated the relative amplitude
$I_2(r)/I_0(r)$ of the $m=2$ of the light distribution along the
radius in real barred-spiral galaxies. Similar estimations of
$I_2(r)/I_0(r)$ are given by Buta, Block \& Knapen (2003) and by
Buta et al. (2005) for different samples. Under the assumption of
constant mass-to-light ratio along the radius of a galaxy, the
quantities $I_2/I_0$ and $A_2(r,t)$ are the same. In the region of
spiral arms the values of $A_2(r,t)$ given above are in quite good
agreement with the values of $I_2/I_0$ found in real galaxies by
the above authors. The same is true for the four models.

In order to estimate the time evolution of the relative
contribution of the $m=2$ mode in the whole power spectrum of
$\delta(r,\phi,t)$ we define the mean value $<a_2^2>(t)$ of
$a_2^2(r,t)$ from all the rings with radii between the corotation
radius $r_c$ and $r_c+1$. In Fig. 10 the quantity $<a_2^2>(t)$ is
plotted as a function of time for all the rotating models. We see
that $<a_2^2>(t)$ oscillates up and down, but on the average it
decays with time, tending to a minimum value somehow smaller than
0.2 at $t_2=300$. It is remarkable that the mean values of
$<a_2^2>(t)$ at early times (e.g. at $t_1 \approx 20 $), as well
as the time intervals when $<a_2^2>(t)$ maintains values greater
than a certain value, e.g. $0.3$, increase along the sequence of
the models.

Notice that the bold dots drawn on the top of six of the peaks in
Fig. 10 indicate the snapshots displayed in Fig. 5. An example of
the evolution between two successive peaks is described below in
Fig. 11.

The oscillations of $<a_2^2>(t)$ in Fig. 10 show a periodicity
related to the period of rotation of the bar. This effect
certainly betrays information related to the mechanism of
excitation and propagation of spiral waves. It requires a much
thorough investigation, but here, we will give a short qualitative
description of what seems to happen, based only on the observation
of a number of successive snapshots. In order to see this effect
in more details, we select a series of 12 successive snapshots (at
every $\Delta t=1 T_{hmct}$) in the time interval between $t=138$
and $t=149$ of the evolution of the QR4 model during the 18th-19th
rotation of the bar. This interval of time is marked by a box with
dotted lines in Fig. 10d. During this interval the quantity
$<a_2^2>(t)$ describes about 1.5 oscillations, i.e.
minimum-maximum-minimum-maximum.

In Fig. 11 the distributions of $\delta_2(r, \phi, t)$ and
$\delta_{3imp}(r,\phi,t)$ are shown (in pairs as in Fig. 5) for
the 12 successive snapshots ($t=138\rightarrow149$). Observing
carefully these distributions we can describe the following
mechanism:

Spiral pattern is due to a wave fluctuation mode of $m=2$ on the
surface density, that is excited by the bar. This wave travels
outwards and rotates more slowly than the bar. The bar goes out of
phase with respect to a previously excited part of the mode. Two
parts of this mode, excited at a small difference of time are not
always in phase. In this case the continuation of the spiral shape
of the wave can be interrupted. The quantity $<a_2^2>(t)$ has a
minimum. The two parts of this mode come in phase
quasi-periodically producing a global spiral wave with maximum
amplitude $<a_2^2>(t)$ and smoothly joined azimuthal phases. They
come out of phase again and the global spiral structure is
temporarily deformed, or interrupted and so on.

It is possible in this case that, temporarily, Quarter-Turn
Spirals (QTS) appear as, for example, in the snapshot at $t=147$
(Fig. 11). Notice that QTS are observed e.g. in NGC 3631 and NGC
1365, (Fridman, Khoruzhii \& Polyachenko 2002; Polyachenko \&
Polyachenko 2002; Polyachenko, 2002a, 2002b).

It is also possible that spiral arms can be so confused, due to an
irregular dependence of $\phi_2(r,t)$ on $r$ that can hardly be
discernible, but the dependence of $\phi_2(r,t)$ on $r$ becomes
smooth again and continuous spiral arms are re-created again in
about half the period of the bar, or in any case in less than a
period of rotation of the bar. This process is repeated all during
the Hubble time in our experiments, when the bar describes a good
number of rotations ranging from 19 rotations in QR1 to 33
rotations in QR4, although the relative power of the $m=2$ mode
with respect to higher modes decreases by a factor of about 2 or
3.

\subsection{Secular evolution and the mass in chaotic motion}
In general the fraction of mass in chaotic motion, the spatial
distribution of this mass, the mean level of the Lyapunov numbers
and their distribution among the chaotic orbits are important
parameters related to the morphology and the secular evolution of
galaxies.

If chaotic orbits have so small Lyapunov numbers that the rate of
their chaotic diffusion in a Hubble time cannot be efficient, or
if chaotic orbits are only a small fraction of the total mass, or
if they are uniformly distributed inside the equipotential surface
of their energy, they cannot drive any secular evolution in a
galaxy.

Secular evolution can be driven within a Hubble time by chaotic
orbits provided that:

i)  Lyapunov times of chaotic orbits do not exceed the Hubble time

ii) the mass in chaotic orbits with such Lyapunov times is a good
fraction of the total mass and

iii)these chaotic orbits occupy a limited part of their available
phase space.

A better understanding of the various features of our models can
be obtained, if we make a distinction between regular and chaotic
orbits. The method of the distinction and the results obtained are
discussed in the next section.

\section{Mass in chaotic versus mass in regular motion}

We use the method introduced in Voglis et al. (2002) to find the
fraction of mass in chaotic motion at various snapshots of our
models and compare the results.

As a measure of the chaotic character of an orbit one can use the
mean logarithmic divergence between neighboring orbits in a finite
time, called Finite Time Lyapunov Characteristic Number ($FT-LCN$)
defined by
\begin{equation}
FT-LCN=\frac{1}{t}\log{\frac{\xi(t)}{\xi(0)}}
\end{equation}
where $\xi(t)$ is the length at time $t$ of the deviation vector
from an orbit under study  and $\xi(0)$ is its initial value.

The deviation vector $\vec{\xi}(t)$ is defined in the
6-dimensional phase space $\vec{\xi}(t)=(dx, dy, dz, dp_x, dp_y,,
dp_z )$ from the differentials of the position and momentum
variables $(x,y,z, p_x, p_y, p_z)$. It is evaluated from the
variational equations, derived by differentiating the equations of
motion. The coefficients of the variational equations are
evaluated from the solutions of the canonical equations of motion
resulting from the Hamiltonian (9), when this Hamiltonian is
applied to a particular snapshot of the self-consistent run of our
systems.

As it is well known (see e.g. Contopoulos 2004, for a review) the
maximal Lyapunov Characteristic Number $LCN$ of an orbit is the
limit of $FT-LCN$ as $t$ tends to infinity. If $LCN$ is zero the
orbit is regular (called also ordered), while if $LCN$ acquires a
positive constant value the orbit is chaotic and the constant
value of the $LCN$ is a measure of the chaotic character of the
orbit.

Since in practice the orbit can be integrated only for a finite
time $t_{max}$, $LCN$ can only be approximated  by the values of
$FT-LCN$. However, the values of $FT-LCN$ never become exactly
zero. In regular orbits the values of $FT-LCN$ tend to zero on the
average as $\log{t}/t$ (with an average slope $1/t$). In chaotic
orbits, for a transient period $t_L$, the values of $FT-LCN$ again
decrease on the average following the same law ($\log{t}/t$), but
later on they tend to be stabilized at a constant value $L$. This
value can be considered as an estimate of the $LCN$ of the orbit.

If $t_L$ is known, a first estimate of $L$ is $L \approx
\log{t_L}/t_L > 1/t_L$. Thus the detectable values of $LCN=L$ of a
chaotic orbit are limited by the inequality $L > 1/t_{max}$. In
order to obtain a distinction between chaotic and regular orbits
in terms of $L$, it is necessary to define a threshold value of
$L=L_{min} \gtrsim  1/t_{max}$. All the orbits for which $FT-LCN$
has already been stabilized  at a value $L > L_{min}$, at
$t=t_{max}$, can be characterized as chaotic. The maximum Lyapunov
time of these orbits is $T_{L, max}=1/L_{min}$. For insuring that
all the orbits, that practically behave as chaotic, have been
detected, $T_{L, max}$ should be remarkably larger than a Hubble
time.

On the other hand, the orbits for which $FT-LCN$ at $t=t_{max}$
still decreases on the average as $\log{t}/t$ could be either
regular (with $LCN=0$), or chaotic but with $LCN < L_{min}$. All
these orbits can conventionally be characterized as regular
orbits.

A difficulty arising in applying this method in galaxies is the
fact that in galaxies there are orbits evolving in very different
characteristic time scales. The radial period $T_{rj}$ (i.e. the
time for a orbit to go from its apocenter to pericenter and back
to apocenter) of the orbit $j$ depends on the binding energy $E_j$
of this orbit according to the relation $T_{rj} \sim \mid E_j \mid
^{-3/2}$, almost independently of the angular momentum of the
orbit. This relation is exact in a Keplerian field, as well as in
the isochrone potential model. In gravitational N-body
simulations, the same relation still holds with a remarkable
accuracy (Voglis, 1994).

In galaxies the binding energies of individual orbits, and
therefore the corresponding radial periods, spread to more than
four orders of magnitude. The departure from integrability of an
orbit depends on the variability per dynamical time of the actions
associated with the orbit. If in an orbit a certain variation of
actions takes place in 1 dynamical time  and in another orbit the
same variation takes place in 10 dynamical times, the departure
from integrability is larger in the former case.

Therefore, a fair estimate of the departure of orbits from
integrability can be obtained by integrating the orbits for the
same number $N_{r,max}$ of radial periods than for the same time
$t_{max}$.

For this reason we use the Specific Finite Time Lyapunov
Characteristic Number ($SFT-LCN$) denoted by $L_j$ for every orbit
$j$. The term `Specific' is used to declare that $L_j$ is
evaluated in units of the inverse radial period $1/T_{rj}$ of the
particular orbit. This makes the values of $L_j$ almost invariant
in the characteristic time scale of the orbit. A particular
threshold value of $L_j=L_{min}$ is a better index for distinction
between regular and chaotic orbits.

A disadvantage of the index $L_j$ is that the Lyapunov times
measured by $T_{L}=T_{rj}/L_j$ (in fixed time units) are different
for orbits with the same $L_j$, but different energies. The
chaotic diffusion of an orbit is important for times greater than
its Lyapunov time $T_L$. Among the chaotic orbits, only those with
$T_L < t_{Hub}$ are able to develop effective chaotic diffusion in
a Hubble time.

In order to estimate the ability of orbits for an effective
chaotic diffusion in a Hubble time we use, instead of the index
$L_j$, the index
\begin{equation}
L_{cu}= L_j \frac{T_{hmr}}{T_{rj}},
\end{equation}
which is the value of $FT-LCN$ measured in a common unit
$1/T_{hmr}$ for all the orbits, where $T_{hmr}$ is the radial
period of an orbit with binding energy $E_{hmr}$. A Hubble time is
$t_{Hub}= 300 T_{hmct}=100 T_{hmr}$, as described in the
introduction. The Lyapunov times of the orbits in common units of
$T_{hmr}$ are $T_{L}=1/L_{cu}$. Therefore, chaotic diffusion of an
orbit $j$ becomes important in a Hubble time, if $T_L \lesssim
t_{Hub}$, or if $L_{cu} \gtrsim 10^{-2}$.

An alternative index that can be combined with the index $L_j$ for
the distinction between regular and chaotic orbits is the so
called Alignment Index ($AI$) defined as
\begin{equation}
AI=\sqrt{1-\left| \frac{\vec{\xi}_1(t)\centerdot \vec{\xi}_2(t)}
{\xi_1(t)\xi_2(t)}\right|}
\end{equation}
where $\vec{\xi}_1(t)$ and $\vec{\xi}_2(t)$ are two different
deviation vectors of the same orbit. This index is a direct
measure of the angle between the two deviation vectors.

The behavior of the deviation vectors in four-dimensional
symplectic maps or in three-dimensional Hamiltonian systems has
been examined in Voglis, Contopoulos \& Efthymiopoulos (1998,1999)
and Voglis et al. (2002). In the case of ordered motion, two
initially arbitrary deviation vectors, after a transient period of
time, become tangent to the surface of the corresponding torus. On
the surface of the torus they oscillate around each other. Thus,
the angle between the two vectors (and hence the alignment index
$AI$) shows no systematic variation, but remains always around a
finite mean value.

On the contrary, if the orbit is chaotic, the directions of the
two deviation vectors tend exponentially to coincide to each other
along the same direction (parallel or antiparallel), which is the
direction of the nearby passing most unstable manifold. Thus, the
index $AI$ tends exponentially to zero.

In Voglis et al. (2002) the alignment index $AI$ was defined as
the smaller between the sum or the difference of the two deviation
vectors (normalized to unity at every time step), depending on
whether they are antiparallel or parallel, respectively. This
index is often called SALI (Smaller ALignment Index, Skokos 2001).
But by the definition given in (17) the distinction between
parallel or antiparallel deviation vector is avoided.

The adopted maximum number of radial periods of the run time
$t_{j,max}$ of the orbit $j$ is $N_{r,max}=t_{j,max}/T_{rj}=1200$
for all the orbits. The corresponding threshold value of $L_j$ is
$L_j=10^{-2.8}$. By evaluating simultaneously the two indices
($AI_j$ and $L_j$) along the orbits of the system we can obtain a
reliable distinction between ordered and chaotic orbits in our
models at a particular snapshot of their self-consistent run.

It should be stressed that by this distinction we only check
whether the positions and velocities of the particles, if they are
considered as initial conditions in the fixed potential of the
particular snapshot, belong to regular or to chaotic orbits. In
general, the same initial conditions in a self-consistent run,
i.e. in a time varying potential, do not lead exactly to the same
orbits. This distinction regards a particular snapshot and informs
us only about the character of the initial conditions of the
orbits in the fixed potential of this snapshot. However, if the
system is in a well established equilibrium (it has no significant
secular evolution) the coefficients of the potential have only a
small noise (less than 1\%). We have shown (Voglis et al. 2002)
that, in this case, if the method of distinction is applied at a
different snapshot, e.g. a time of $100T_{hmct}$ later, only a
small number of particles in the system (less than $3\%$) have
changed the character of their motion (from regular to chaotic, or
vice versa), because of the noise in the coefficients of the
potential. Thus, an uncertainty in estimating the fractions of
mass in regular or chaotic motion of the order of about $3\%$
always exists.

In a system that develops significant secular evolution, as it is
the case of our rotating systems, or the case when the system
contains a massive central concentration (e.g. Kalapotharakos et
al. 2004), the coefficients of the potential can considerably vary
in time, resulting in a considerable change on both the number of
chaotic orbits and on the distribution of the Lyapunov numbers.
Useful pieces of information about the secular evolution of such
systems can be obtained by repeating the distinction between
regular and chaotic orbits at several snapshots.

In Fig. 12a the fractions of mass detected in chaotic motion are
shown by points at the snapshots of $t_1 \approx 20 $ and
$t_2=300$ joined by lines for all the models as indicated in the
figure. In the non-rotating model Q the fraction of mass in
chaotic motion remains constant at about $32\%$ all along the
period of a Hubble time. In the rotating models the fraction of
mass in chaotic motion is at the level of about $60\%$ to $65\%$
and remains on this level, presenting only a small decrease (of a
few \%) at the end of the Hubble time. (Notice that all the
fractions of mass in this article are with respect to the total
mass of the system).

A similar representation is shown in Fig. 12b, but for the
fractions of mass with $L_{cu}>10^{-2}$, that can develop chaotic
diffusion in a Hubble time. These fractions range from about
$45\%$ to $55\%$ at $t_1$ and from about $35\%$ to $45\%$ at
$t_2$. In other words, the Lyapunov numbers $L_{cu}$ of a fraction
of about $10\%$ of the total mass have been shifted to values
smaller than $10^{-2}$ at $t_2$. This indicates a tendency of the
mass to be self-organized to some degree.

The distributions of chaotic orbits along the axis of
$\log{L_{cu}}$ are shown in Fig. 13 for the four rotating models.
In every model the solid line comes from the snapshot at $t_1$ and
the dashed line comes from the snapshot at $t_2$. The dotted line
with index Q in  this figure gives the distribution of the Q
model, for comparison. All the distributions are normalized with
respect to the total mass of the system.

At $t_1$ the maximum of these distributions occurs at values of
$\log{L_{cu}}$  about $-1.7, -1.5, -1.5, -1.7$ for QR1, QR2, QR3,
QR4, respectively, i.e. by about one order of magnitude larger
than in the Q model, in which the maximum of the distribution
occurs at about $-2.5$. This difference is important for the
chaotic diffusion in a Hubble time. A direct consequence of this
shift of the Lyapunov numbers is that the fraction of mass that
can develop effective chaotic diffusion in a Hubble time ($L_{cu}
> 10^{-2}$) presents a considerable increase from the level of
$8\%$ to the level of $45\%-55\%$, as mentioned above.

It comes out, therefore, that rotation enhances chaos, not only as
regards the fraction of mass in chaotic motion, but also as
regards the magnitude of the Lyapunov numbers, that are seriously
shifted towards larger values. This impressive enhancement of
chaos is due to the rotation.

Rotation produces two effects. A re-distribution of mass that
creates a different self-consistent potential and a series of
resonances that affect the dynamical evolution of the system. A
reasonable question at this point is, which of the two effects is
more responsible for the enhancement of chaos?

As we have seen in the models Q and C, if there is no rotation
chaos increases with increasing departure from sphericity, e.g.
with increasing maximum ellipticity. The new distribution of mass
caused by rotation is more complicated than the distribution of
mass in the non-rotating models. Namely, the whole distribution of
mass is flatter, but the density becomes larger near the center. A
central bar is formed surrounded by a thin and a thick disc. On
the disc the surface density shows an exponential radial
dependence and a spiral azimuthal dependence. This more
complicated structure is expected to be in favor of chaos.

On the other hand, the rotation of the above pattern with a speed
$\Omega_p$ introduces instabilities and a sequence of important
overlapping resonances between $\Omega_p$ and the frequencies of
orbits. Interactions between resonances is known to be an
important source of chaos (Contopoulos 1966; Rosenbluth et al.
1966; Walker \& Ford 1969; Chirikov, Israilev \& Shepelyansky
1971; Zaslavsky \& Chirikov 1972; Chirikov 1979. For a review see
Contopoulos 2004).

These two effects cannot be disentangled. However, we can check to
what extend each effect is responsible for chaotic motion by the
following test. We consider, for example, that the distribution of
mass at a given snapshot, e.g. at $t_1\approx 20$, of QR3 does not
rotate. We run the orbits in the fixed potential of this snapshot
without rotation ($\Omega_p=0$) and separate the mass in regular
and in chaotic motion by the same method. The resulting
distribution of the chaotic mass along the $\log{L_{cu}}$ axis in
this case  (non-rotating potential of QR3 at $t_1$) is shown by
the dotted line with index $t_1(\Omega_p=0)$ in Fig. 13c. The
corresponding fraction of mass is about 46\% of the total mass,
while the fraction with $\log{L_{cu}}> -2$ is  only about 15\%.
Notice that, more than 95\% of the identities of particles under
the curve $t_1(\Omega_p=0)$ in Fig. 13c appear also in the group
of particles under the curve $t_1$ of the same figure.

The enhancement of chaos between the curves Q and
$t_1(\Omega_p=0)$ can be attributed to the scattering of high
energy orbits by the bar and to the torques due to the spiral
structure. Such torques are not negligible. (Notice that this is
also true in real galaxies, see e.g. Gnedin, Goodman \& Frei 1995;
Kormendy \& Kennicutt 2004; Block et al. 2004). Thus, we conclude
that the complicated pattern itself (formed by the rotation)
increases the mass in chaotic motion even under the absence of
resonances.

In the rotating pattern, resonances increase further the fraction
of chaotic orbits. Furthermore, instabilities introduced by the
resonances and interactions between them produce a remarkable
increase on the mean level of the Lyapunov numbers. (Compare the
curves $t_1(\Omega_p=0)$ and $t_1$ in Fig. 13c).

Thus, we conclude that rotation enhances the mass in chaotic
motion, as well as the mean level of the Lyapunov numbers by two
collaborating mechanisms, i.e. the more complicated distribution
of mass and the instabilities introduced by resonances and
interactions between resonances.

During the secular evolution of the systems, part of the mass in
chaotic motion is self-organized, or tends to smaller Lyapunov
numbers. Comparing the distributions of the chaotic mass along the
$\log{L_{cu}}$ axis at $t_1 \simeq 20$ (solid line) and $t_2=300$
(dashed line) in Fig. 13, shows that the maximum is shifted
towards smaller values and the mass in chaotic motion decreases.
The mass with values of $\log{L_{cu}}$ above the limit of $-2$ at
$t_2=300$ becomes by about 10\% smaller as described above (Fig.
12).

A question at this point is how the self-organization observed in
this investigation can be compatible with the second law of
thermodynamics. What happens to the entropy?

Self-gravitating systems have negative specific heat (Lynden-Bell
\& Wood 1968; Binney \& Tremaine 1987). The angular momentum
transference outwards causes an expansion of the outer material of
the disc and a shrinking of the inner material. The energy is
redistributed along the radius of the system. In the inner parts
the energy decreases, while in the outer parts the energy
increases. An example of comparison of the binding energies of
individual particles from QR4 is shown in Fig. 14 ($E(t_1)$ at
$t_1$ and $E(t_2)$ at $t_2$). Particles at lower energies lose
some of their energy and are projected below the diagonal in this
figure, while particles with high energies gain energy and they
are projected above the diagonal. Some particles can acquire even
positive energies and escape from the system, so the energy of the
bound part becomes smaller. This really happens for about $0.5\%$
of the particles in each model, as already mentioned at the end of
Section 2. The redistribution of energies is associated with a
redistribution of entropy. Some entropy of the inner parts is
transferred outwards. Thus, although the total entropy of the
system is expected to increase according to the second law of
thermodynamics, self-organization is possible due to the
redistribution of the binding energies that allows redistribution
of the entropy as well.

\section{Spatial distribution of the mass in chaotic motion}

The values of the Jacobi integral of the majority of regular
orbits belong to the deepest parts of the potential well, while
the majority of chaotic orbits have values of the Jacobi integral
around the maximum value $h_J(r_c)$ of the effective potential at
the corotation radius $r_c$.

In Fig. 15 the effective potential in the rotating frame (right
vertical axes V in each panel) is given as a function of the
radius $r$ (lower horizontal axis r), when $r$ is taken along the
longest axis of the bar (dotted line). (The values of the
potential are normalized so that the deepest value of the
potential at the center of QR1 is $-100$). A horizontal straight
line is also plotted at the maximum value $h_J(r_c)$ of this
potential.

In the same figures, for comparison, the left vertical axis gives
the values of the Jacobi integral $h_J$ of the orbits. The upper
horizontal axis gives the fraction of mass $\Delta N_c(h_J)/N$ of
chaotic orbits (dashed curve) and the fraction of mass in regular
orbits $\Delta N_r(h_J) /N$ (solid curve) as they are distributed
along the $h_J$ axis.

The left column of Fig. 15 refers to the four models at $t_1$. The
right column is similar to the left column, but at $t_2$.

As it can be seen in Fig. 15 chaotic orbits are concentrated
around the maximum values $h_J(r_c)$ of the effective potential
(the unstable Lagrangian points $L1$, $L2$). The maximum of the
distribution of chaotic orbits always occurs for energies slightly
above the corotation energy $h_J(r_c)$. Regular orbits dominate in
the inner parts well inside the corotation radius.

The different distribution of regular and chaotic orbits along the
energy axis implies an important difference in their spatial
distribution.

In Fig. 16 the projections of the rotating models on their
equatorial plane are shown at $t=t_1$. The panels of the same row
correspond to the same model at the same time. In this figure a
sample of 1 particle every 12 particles, uniformly distributed
along the mass of the system, is plotted in the left panel for
each model.

From this sample the set of particles that are detected in chaotic
motion at $t_1$ is denoted by C1 and is plotted separately in the
right panel of the same row, while the set of the particles that
are detected in regular motion at this time is denoted by R1 and
is plotted separately in the middle panel.

By comparing the set R1 (middle panel) with the set C1 (right
panel) in a row of these figures we infer that the spiral arms
that are clearly-sighted in QR2, QR3, QR4 are almost completely
formed by chaotic orbits.

This is a striking new result revealing that a mechanism able to
produce spiral arms should be sought in the properties of chaotic
orbits rather than in the properties of regular orbits, at least
in the case of large deviations from axial symmetry (strongly
perturbed disc). Such a mechanism is proposed in Voglis, Tsoutsis
\& Efthymiopoulos (2006). In this article we show that the spiral
structure in these models is guided by the projection on the
equatorial plane of the unstable manifolds emanating from the
unstable periodic orbits, mainly the unstable manifold emanating
from the lagrangian points $L1$, $L2$ (main manifold). This
manifold determines the basic features of the geometry of all the
unstable manifolds in the same chaotic domain. All the unstable
manifolds create long-living soliton-like phase correlations
between orbits (as those described in Voglis 2003). An example of
how the main manifold reproduces the whole grand design pattern in
the model QR3 is given.

During the evolution of the models from $t_1 \approx 20$ to
$t_2=t_{Hub}=300$, most of the particles in R1 or in C1 maintain
their character of motion, but some particles change their
character of motion from regular to chaotic and some other
particles organize their motion from chaotic to regular. Thus, out
of the same initial sample, two new sets of particles in regular
and in chaotic motion are detected at $t_2$. These sets are
denoted by R2 and C2 and they are shown in Fig. 17, which is
similar to Fig. 16, but for the time $t_2$. Spiral arms are not
easily discernible in this figures, although we know that even at
so late times, as $t_2$, the $m=2$ mode has a spiral form at least
in the region between the corotation radius and $r \approx 3 $
(see Fig. 8). This is because the contrast of the surface density
along the spiral arms is so small that cannot be easily recognized
simply by projecting a sample of particles on the equatorial
plane.

It should be noted that the vast majority (92\%-93\%) of the
identities of particles preserve their initial character of motion
(regular or chaotic) up to the end of the Hubble time. They are
common in R1 and R2 or in C1 and C2. About $2\%$-$3\%$ of the
initially regular orbits become chaotic, and about 5\% of the
initially chaotic orbits become regular. Also a number of chaotic
orbits reduce their Lyapunov numbers during a Hubble time, showing
a tendency of self-organization in the system, as discussed above.

The bar in our simulations is formed by the vast majority of
regular orbits (about 30\%-35\%) that remain the same by a factor
more that 90\% all during the run. This regular part of the bar is
surrounded by a layer of chaotic particles (of mass about
10\%-15\%) that belongs also to the bar (chaotic shell of the
bar). There is a continuous exchange of chaotic particles between
the bar and the disc. Part of the orbits in the chaotic shell of
the bar travel outside the bar (beyond the corotation radius)
along the spiral arms or the disc, but a somehow larger number of
other chaotic orbits come into the chaotic shell of the bar and
they are trapped there, as chaotic (or they organize their
motion), so the size of the bar increases in time.

\section{Conclusions and discussion}

We have presented the results of several N-body simulations of
iso-energetic models of galaxies, having different amounts of
angular momenta, but the same scalar virial equilibrium. We have
examined a number of morphological and dynamical properties of
these models and we have found the sets of particles that move in
regular or in chaotic orbits in every model. We have compared the
spatial distribution of these two sets. Our main conclusions can
be summarized as follows:

1) Density waves form a rotating central bar embedded in a
rotating thin and a thick disc. The size of the bar decreases and
the density in the region of the bar increases with increasing
angular momentum of the models.

2) Beyond the ends of the bar an exponential radial profile of the
surface density is formed, resembling the observed corresponding
profiles of disc galaxies.

3) Spiral arms are spontaneously formed on the disc, due to an
$m=2$ spiral wave fluctuation mode on the surface density, excited
by the bar. Spiral arms show an intermittent behavior. They are
deformed, fade or disappear temporarily, but they grow again, in a
time scale less than the rotation period of the bar. The $m=2$
mode preserves its spiral structure outside corotation for a
Hubble time, during which the bar describes about 20 to 30
rotations (depending on the angular momentum of the model). The
relative amplitude of this mode with respect to higher modes and
noise decreases in time. However, in the region of spiral arms the
amplitude of $A_2(r,t)$ at the end of a Hubble time is on the same
level as it is given from observational data.

4) The dynamical rotational velocity curve shows a relatively
small variability in a Hubble time indicating a slow
redistribution of mass. However, the rotational velocity curve,
evaluated from the mean angular momentum of particles inside
successive cylindrical rings around the rotation axis, shows a
remarkable evolution in time. Namely, this rotation curve
decreases in the inner parts, particularly inside one half mass
radius, due to the transference of angular momentum outwards.

5) In faster rotating models (as in QR2, QR3, QR4) the central bar
loses angular momentum and slows down. The pattern speed
$\Omega_p$ of the bar decreases in time and the respective
corotation radius increases by about $50\%$ at the end of a Hubble
time. This result is in good agreement with observational data
comparing the linear sizes of young and old bars.

6) Rotation enhances chaos substantially, both by increasing the
fraction of mass in chaotic motion and by shifting the Lyapunov
numbers to larger values (by about one order of magnitude with
respect to the non-rotating model). Two effects collaborate for
this enhancement of chaos, both due to rotation. The deformation
of the distribution of mass relative to the non-rotating model and
the resonant effects, particularly near corotation. The mass that
can develop effective chaotic diffusion in a Hubble time increases
significantly. In our examples, this mass rises (from the level of
8\% of the total mass in the non-rotating Q model) to the level of
45\% - 55\% in the rotating models.

7) The energies of chaotic orbits are concentrated around the
value of the energy near the corotation radius (based on the
pattern of the central bar), while the energies of regular orbits
are mainly in the deepest part of the potential well.

8) A most remarkable result in this work is that spiral arms are
almost completely formed by mass in chaotic motion. This is
clearly demonstrated by examining the spatial distribution of the
mass in regular and in chaotic motion.

It is significant that in all our models the $m=2$ mode preserves
a spiral pattern up to the end of a Hubble time. This mode is
gradually obscured by excitation of higher modes and noise. At
least part of the power of the higher modes and the noise is
expected to be a real effect, but numerical effects are expected
to be present as well. We should keep in mind that integrating
orbits in a time varying potential is a demanding problem, mainly
as regards the accuracy of reproducing the self-consistent
potential and its evolution.

In the level of the available accuracy in the dissipationless
simulations presented in this paper, our investigation shows only
what gravity, combined with rotation, can do. The relative
strength of the $m=2$ wave could be better preserved  by the end
of a Hubble time under the presence of dissipation. Dissipation
could absorb noisy components and reinforce the robustness of the
$m=2$ mode. Furthermore, gaseous material could be sunk and
trapped well inside the corotation radius, as e.g. in Bournaud \&
Combes (2002); Kormendy \& Kennicutt (2004). This process enriches
the bar with angular momentum, competing the angular momentum
transference outwards.

Other parameters, not involved in this study, could play a role as
well. E.g. the presence of a massive central black hole could
moderate the strength of the bar (or even destroy the bar, see
e.g. Shen \& Sellwood 2004). Furthermore, the rate of
redistribution of angular momentum can be modified by the presence
of a halo with low values of $\lambda$. However, chaos can still
play an important role in the dynamics and morphology of these
systems.

\bigskip

\section*{Acknowledgments}

We are grateful to the referee D. A. Gadotti for his remarks,
questions and comments that led us to investigate more deeply
several points and make a substantial improvement of the paper. We
wish to thank Drs. A. Allen ,P. Palmer and J. Papaloizou for their
code. We wish also to thank Prof. G. Contopoulos, Dr. Ch.
Efthymiopoulos and Dr. P. Patsis for useful discussions and
comments. I.S. wishes to thank the Greek State Scholarship
Foundation (I.K.Y) for financial support. This research was also
partly supported by a research program of the EMPEIRIKEION
Foundation.

\clearpage

\begin{figure*}
\centerline{\includegraphics[width=\textwidth]{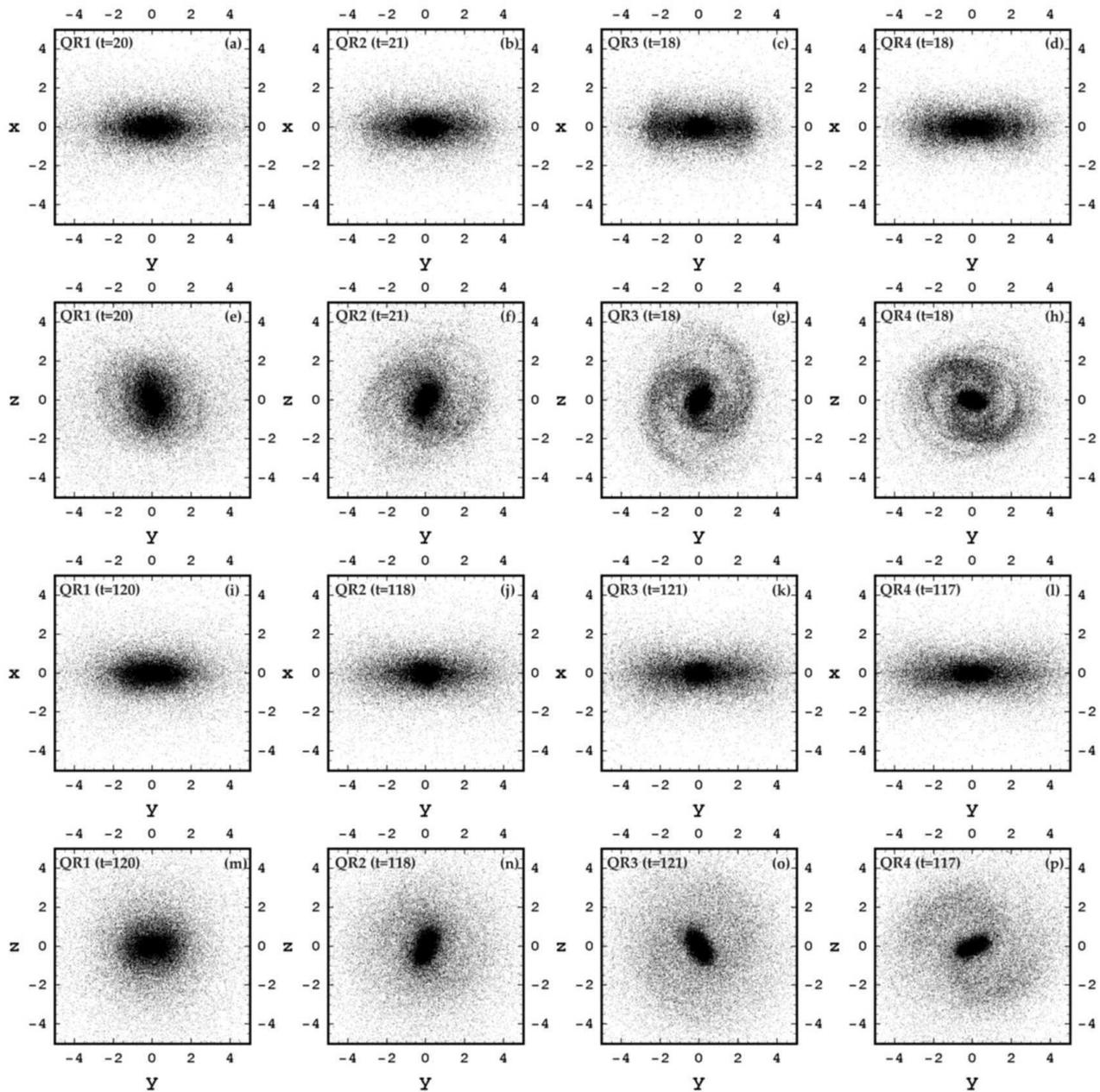}}
\caption{\textbf{First row:} Edge-on projections of the rotating
models at $t_1=20, 21, 18, 18$. \textbf{Second row:} Face-on
projection at the same snapshots as in the first row. Density
waves form a rotating bar embedded in a thick disc. The size of
the bar decreases with increasing angular momentum along the
sequence of the models. Spiral arms are excited on the disc that
are more clearly-sighted in the models QR2, QR3, QR4.
\textbf{Third row:} As in the first row but at a time about
100$T_{hmct}$ later. The total number of rotations from $t=0
\rightarrow t\approx 120$ is 9, 11, 13, 15 for the four models
respectively. An expansion on the plane of the disc has occurred,
increasing along the sequence of the models. \textbf{Fourth row:}
Face-on projections of the same snapshots as in the third row. In
QR4 spiral arms are still well discernible. In QR2, QR3 they are
fainter, and in QR1 cannot be seen in this projection (see Section
3.5). As it is shown in Section 5 (Fig. 16, Fig. 17) these spiral
arms are almost completely formed by chaotic orbits.} \label{fig1}
\end{figure*}
\clearpage

\begin{figure*}
\centerline{\includegraphics[width=30pc]{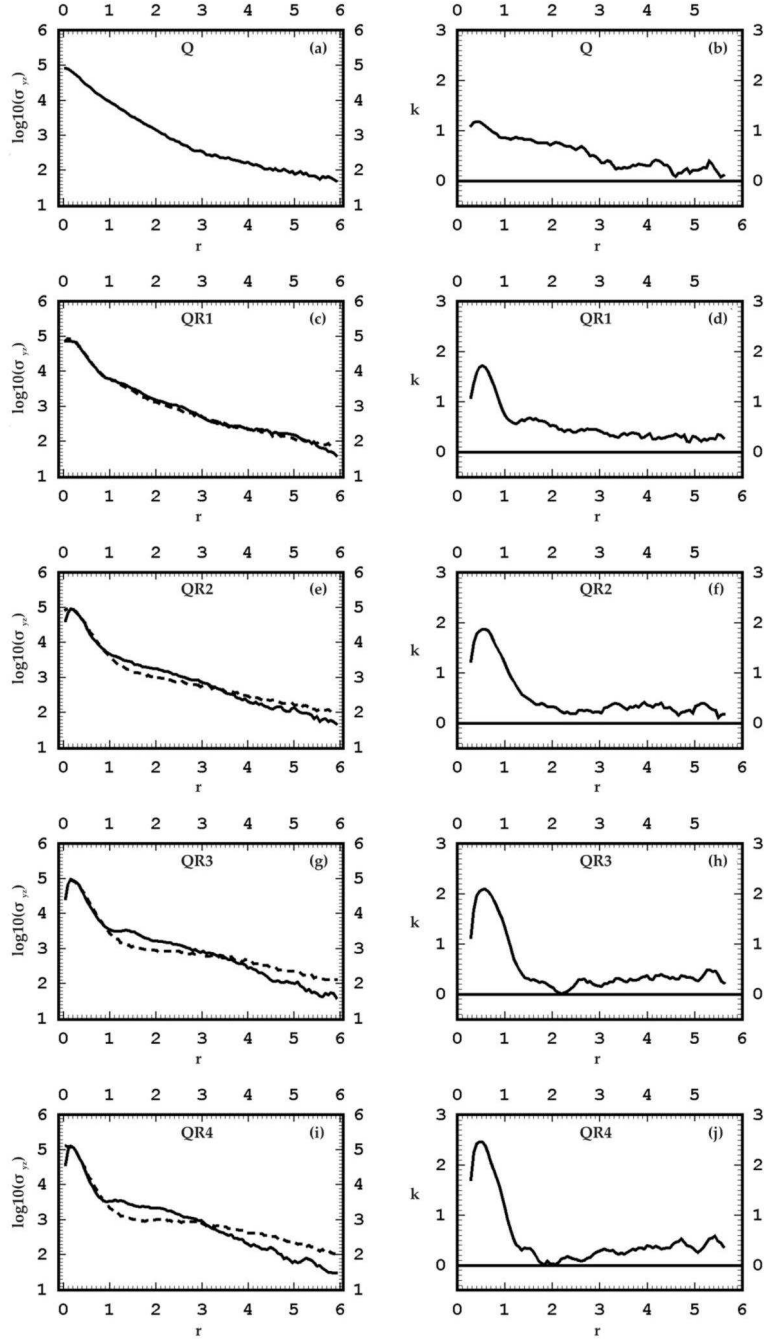}}
\caption{\textbf{Left column:} Radial profiles of the mean surface
density $\overline{\sigma}_{yz}(r)$ for all the models at $t_1
\approx 20$ (solid line) and at $t_2=300$ (dashed line). At small
radii $r \lesssim 1 $ the profiles become steeper and steeper
along the sequence of the models. At larger radii the profiles in
the rotating models become exponential. The exponential profiles
show a slow time evolution tending to be flatter. (Solid line is
at $t_1 \thickapprox 20$, while dashed line is at $t_2=300$).
\textbf{Right column:} The slope $k$ of the profiles as a function
of the radius at $t=150$. In Q this slope decreases gradually (on
the average) along the radius. In the rotating models $k$ has a
pronounced maximum at $r < 1$, but falls to a smaller roughly
constant value at larger radii.} \label{fig2}
\end{figure*}
\clearpage

\begin{figure*}
\centerline{\includegraphics[width=30pc]{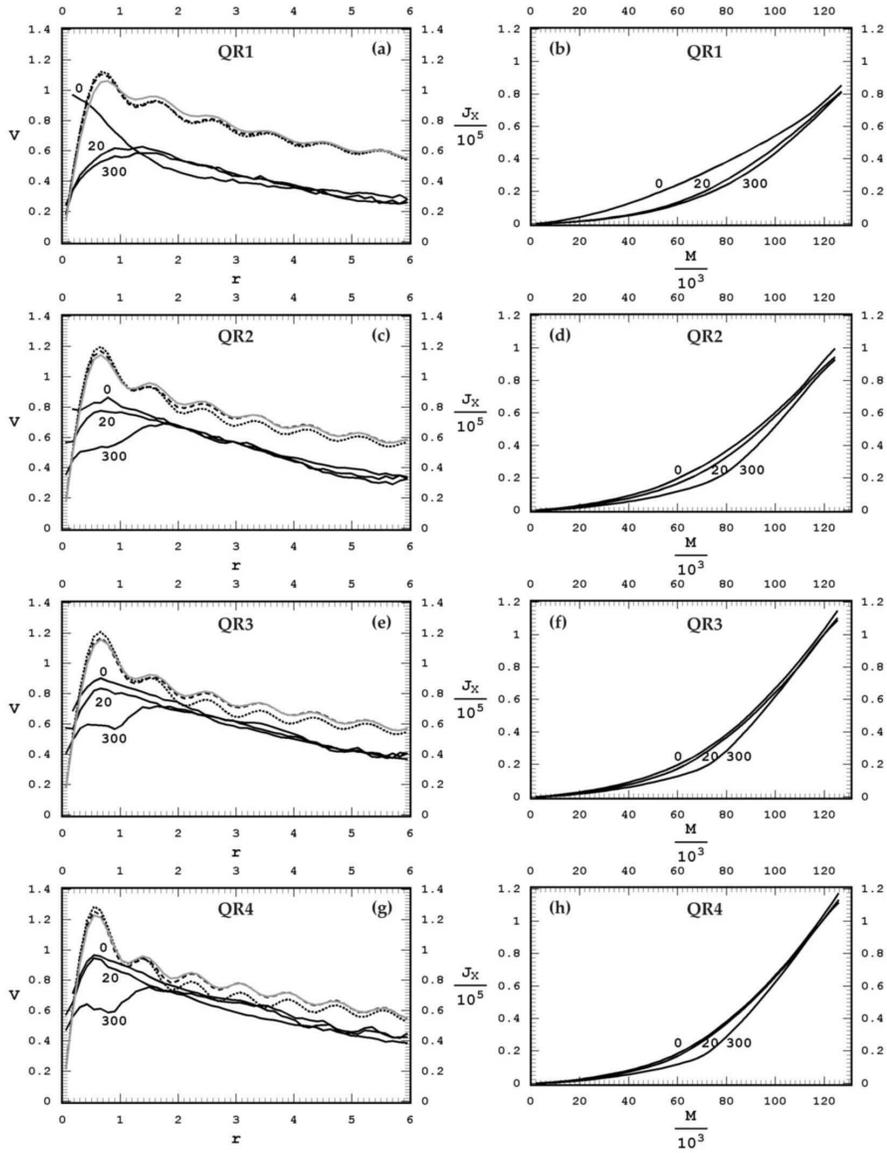}}
\caption{\textbf{Left column:} Rotational velocity curves. Solid
lines give the rotational velocity $v_i(r_i)$ evaluated from the
mean angular momentum per particle in successive cylindrical rings
at  times $t=0, 20, 300$ as indicated by the numbers in the
figures. These curves show that $v_i(r_i)$ decreases in time
particularly in the inner parts, due to the transference of
angular momentum outwards. The dashed lines give the ``dynamical"
rotational velocity curves evaluated from the monopole terms of
the potential at $t_1 \approx 20$ and $t_2=300$. These curves show
a small variation during this period, due to a slow redistribution
of mass. \textbf{Right column:} The cumulative angular momentum vs
the cumulative mass from the center at the same times as in the
left column. At later times more mass is required in order to
collect the same amount of angular momentum, as a result of the
transference of the angular momentum outwards.}\label{fig3}
\end{figure*}
\clearpage

\begin{figure*}
\centerline{\includegraphics[width=\textwidth]{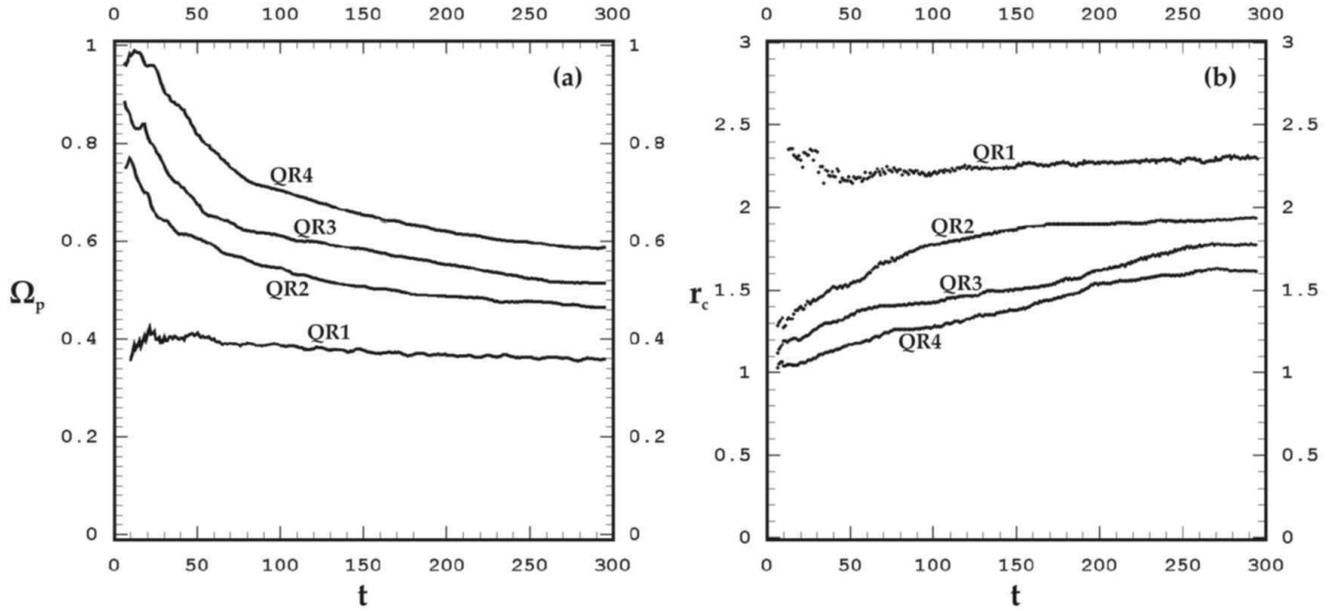}}
\caption{\textbf{(a)} Time evolution of the pattern speed
$\Omega_p$ of the bar in every model. \textbf{(b)} Time evolution
of the corotation radius $r_c$ (radius of the unstable Lagrangian
point L1). In QR1 the pattern speed and the corotation radius
remain roughly constant. In QR2, QR3, QR4 $\Omega_p$ decreases to
about 60\% the initial value, while the corotation radius increase
to about 150\% the initial value.}\label{fig4}
\end{figure*}
\clearpage

\begin{figure*}
\centerline{\includegraphics[width=14.0cm]{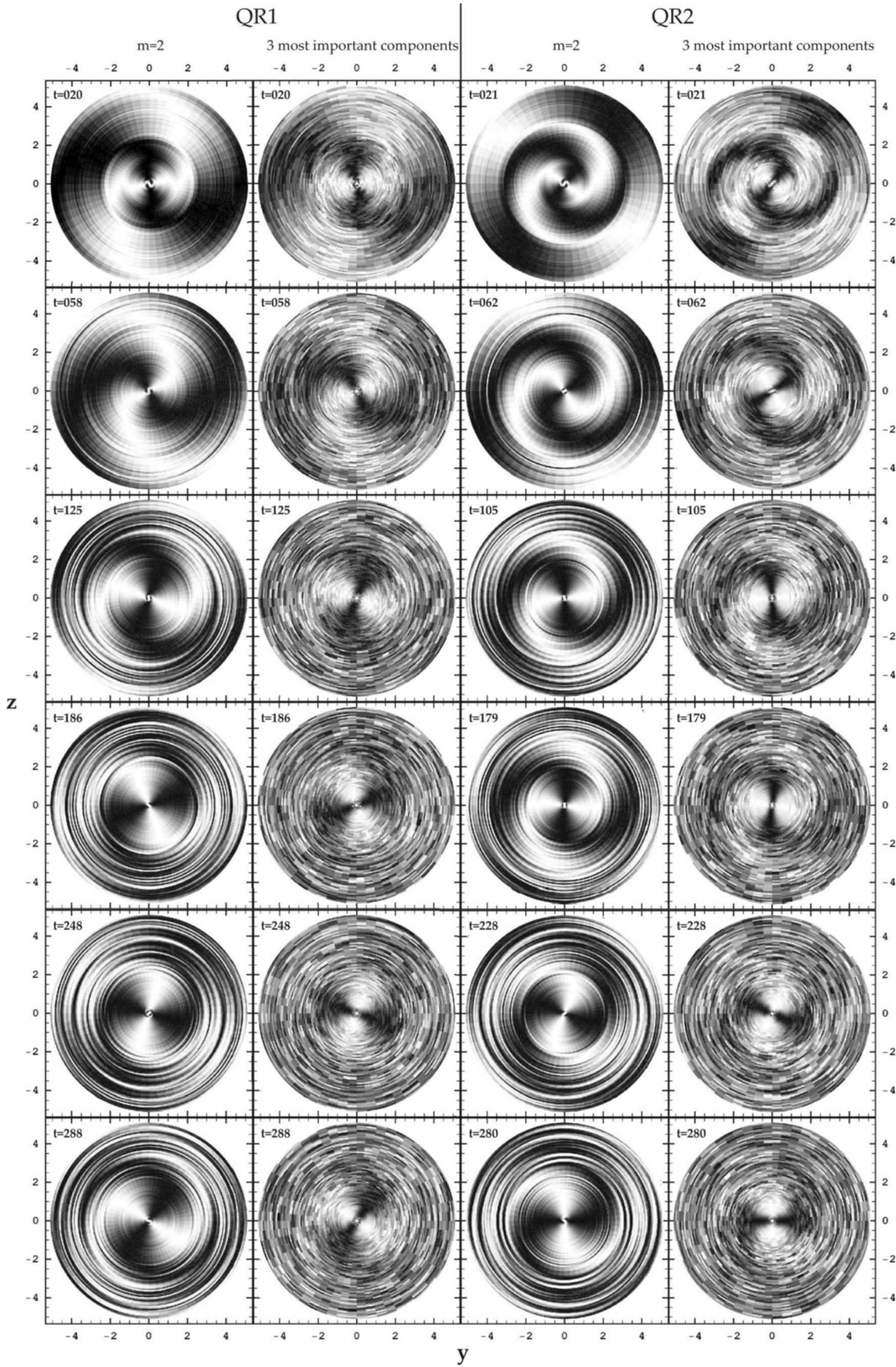}}
\caption{Two columns are plotted for every rotating model. For
each model the \textbf{left column} gives six snapshots of the
distribution on the equatorial plane $(r, \phi))$ of the function
$\delta_2 (r, \phi, t)$, i.e. the mode $m=2$ of the fluctuations
of the surface density. This mode forms a spiral pattern starting
slightly inside the corotation radius and going out beyond the
corotation to large radii in all these snapshots. The
\textbf{right column} gives the corresponding distributions of
$\delta_{3imp} (r, \phi, t)$, resulting from the superposition of
the three most important modes of the spectrum. The $m=2$ mode is
obscured to some extent by other modes, but it can be discernible
in the distributions of $\delta_{3imp} (r, \phi, t)$ even at the
latest snapshots of the simulation, as in the bottom row in this
figure.}\label{fig5}
\end{figure*}
\clearpage

\addtocounter{figure}{-1}

\begin{figure*}
\centerline{\includegraphics[width=14cm]{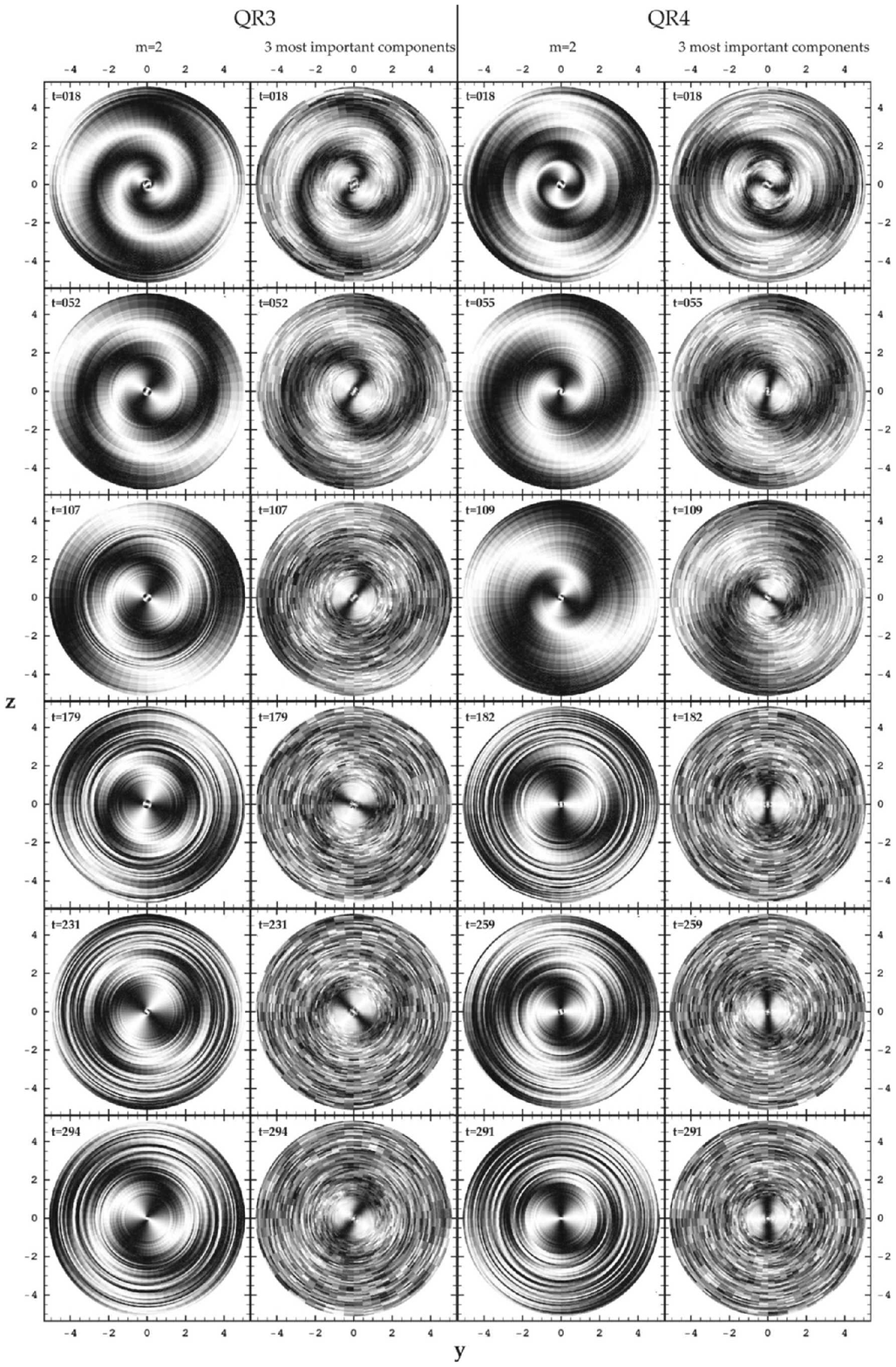}}
\caption{(Continued).}\label{fig5}
\end{figure*}
\clearpage

\begin{figure}
\centerline{\includegraphics[width=8.5cm]{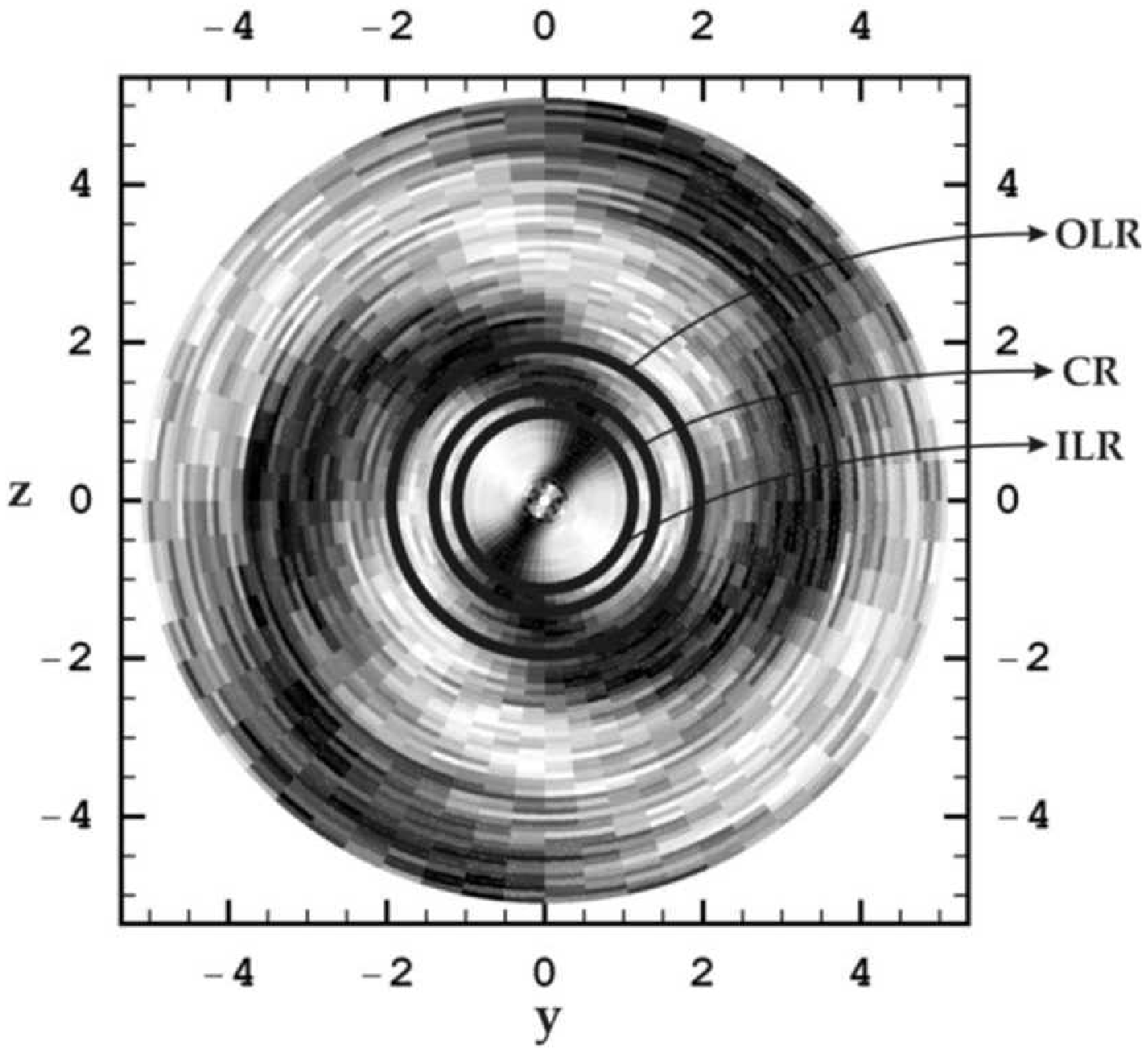}}
\caption{A magnification the distribution of $\delta_2 (r, \phi,
t)$ at the snapshot of QR4 at $t=55$ (second row, seventh column
of Fig. 5), where the three cycles at the radii of the resonances
ILR, CR, OLR are drawn.}\label{fig6}
\end{figure}
\clearpage

\begin{figure*}
\centerline{\includegraphics[width=\textwidth]{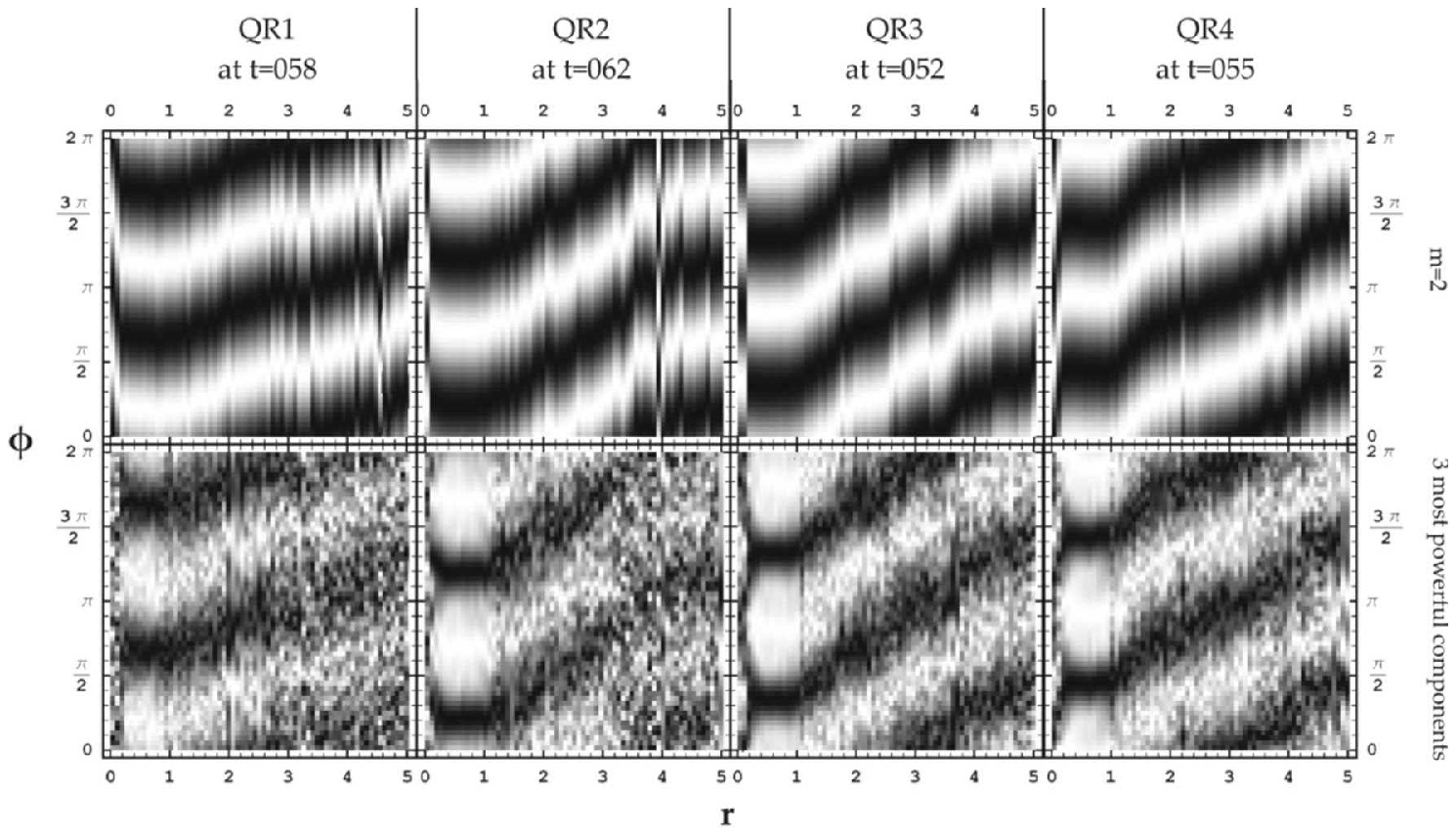}}
\caption{\textbf{First row:} Distributions of $\delta_2 (r, \phi,
t)$  on a rectangular frame $(r, \phi)$ for the four models at the
snapshots of the second row of Fig. 5. Beyond the corotation
radius the phase $\phi_2(r,t)$ is (on the average) a monotonic
almost linear function of $r$, revealing a spiral pattern.
\textbf{Second row:} As in the first row, but for $\delta_{3imp}
(r, \phi, t)$. The obscuration of $\phi_2(r,t)$ by the other modes
is rather small.}\label{fig7}
\end{figure*}
\clearpage

\begin{figure*}
\centerline{\includegraphics[width=\textwidth]{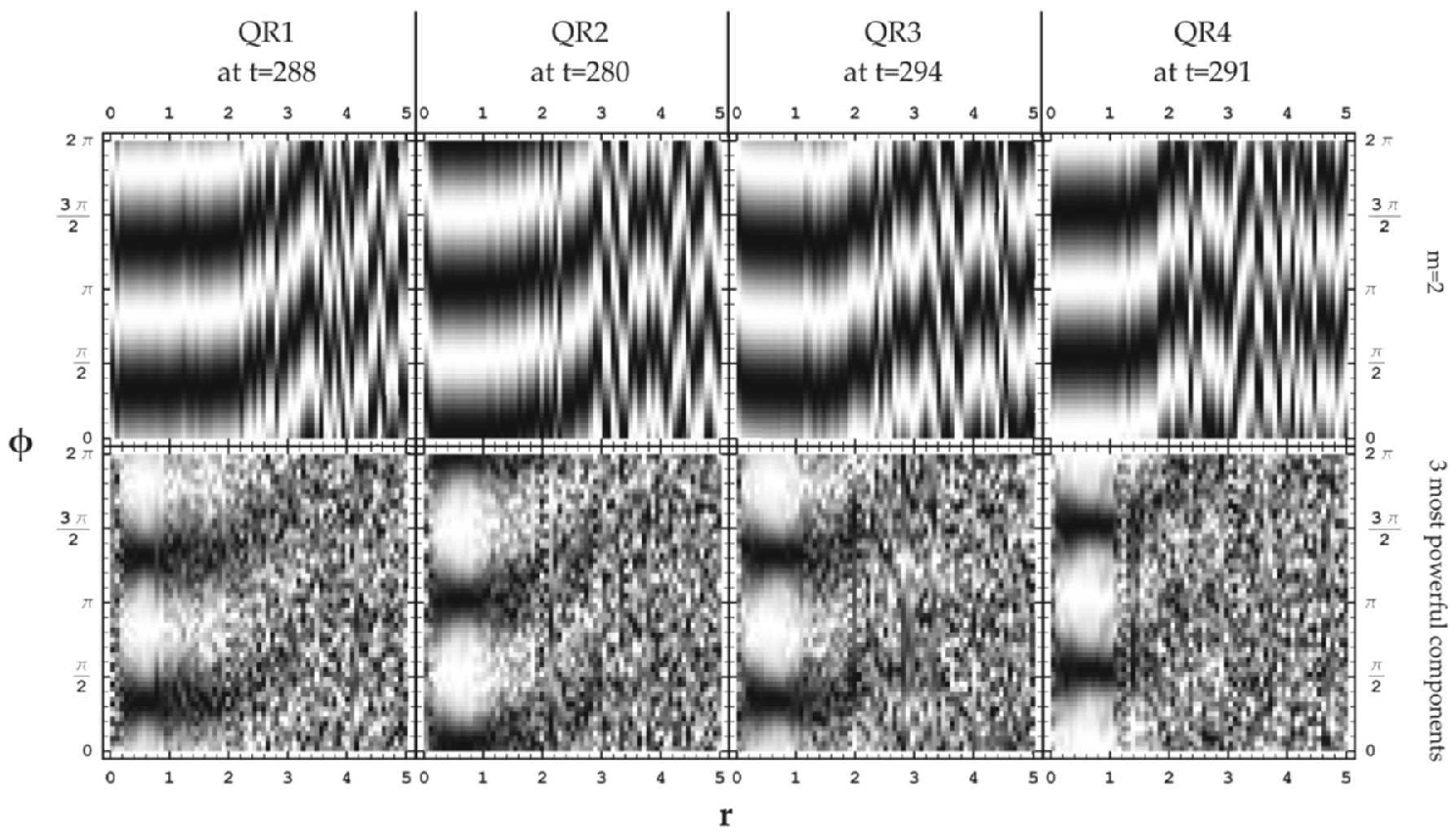}}
\caption{As in the first and the second row of Fig. 7, but for the
snapshots of the last row in Fig. 5. In the first row the dark
lanes are along a straight line segment between the corotation
radius and $r \approx 3$ or $3.5$. In the second row, despite the
obscuration by other modes at this late snapshot, the spiral
structure of the $m=2$ mode is still discernible.}\label{fig8}
\end{figure*}
\clearpage

\begin{figure*}
\centerline{\includegraphics[width=\textwidth]{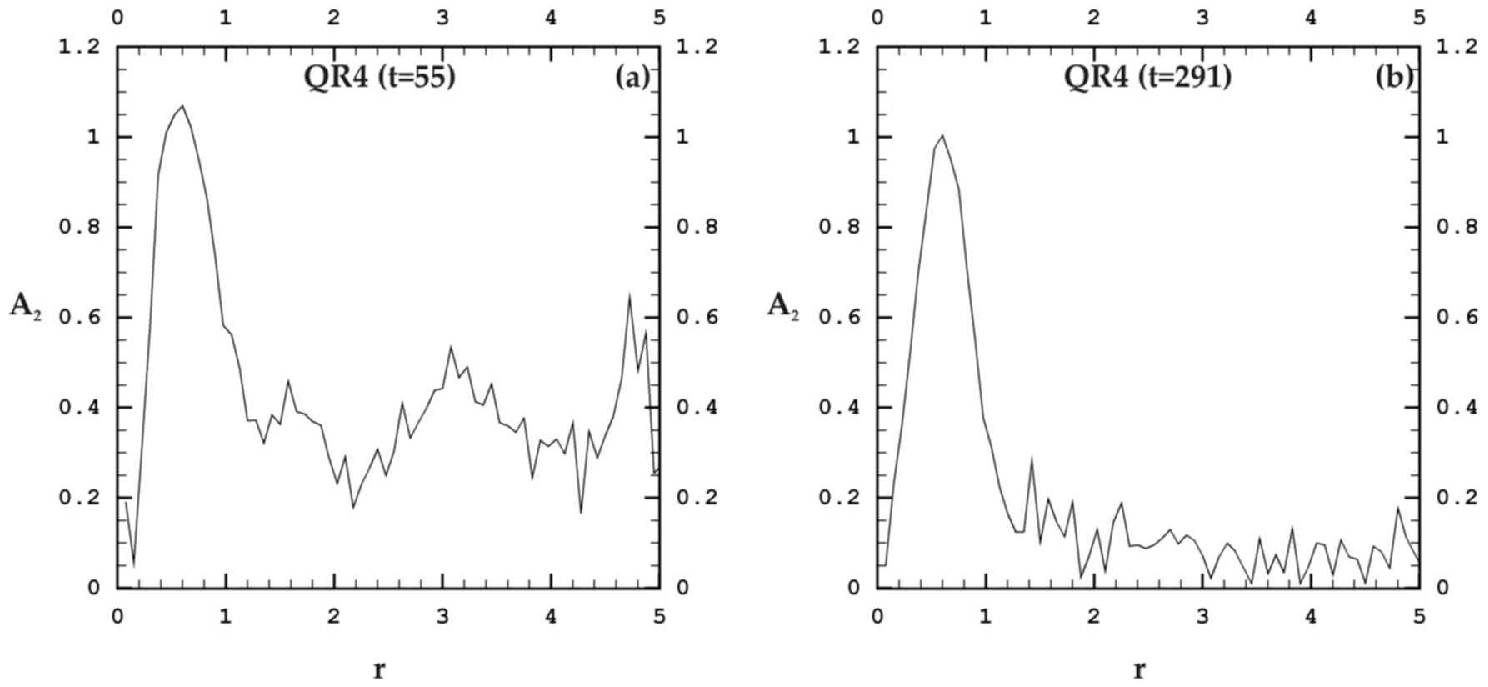}}
\caption{The size of the amplitude $A_2(r,t)$ of the $m=2$ mode
(eq. 14) as a function of the radius $r$ at the two snapshots of
QR4 shown in the second and the sixth row of Fig. 5.}\label{fig9}
\end{figure*}
\clearpage

\begin{figure*}
\centerline{\includegraphics[width=\textwidth]{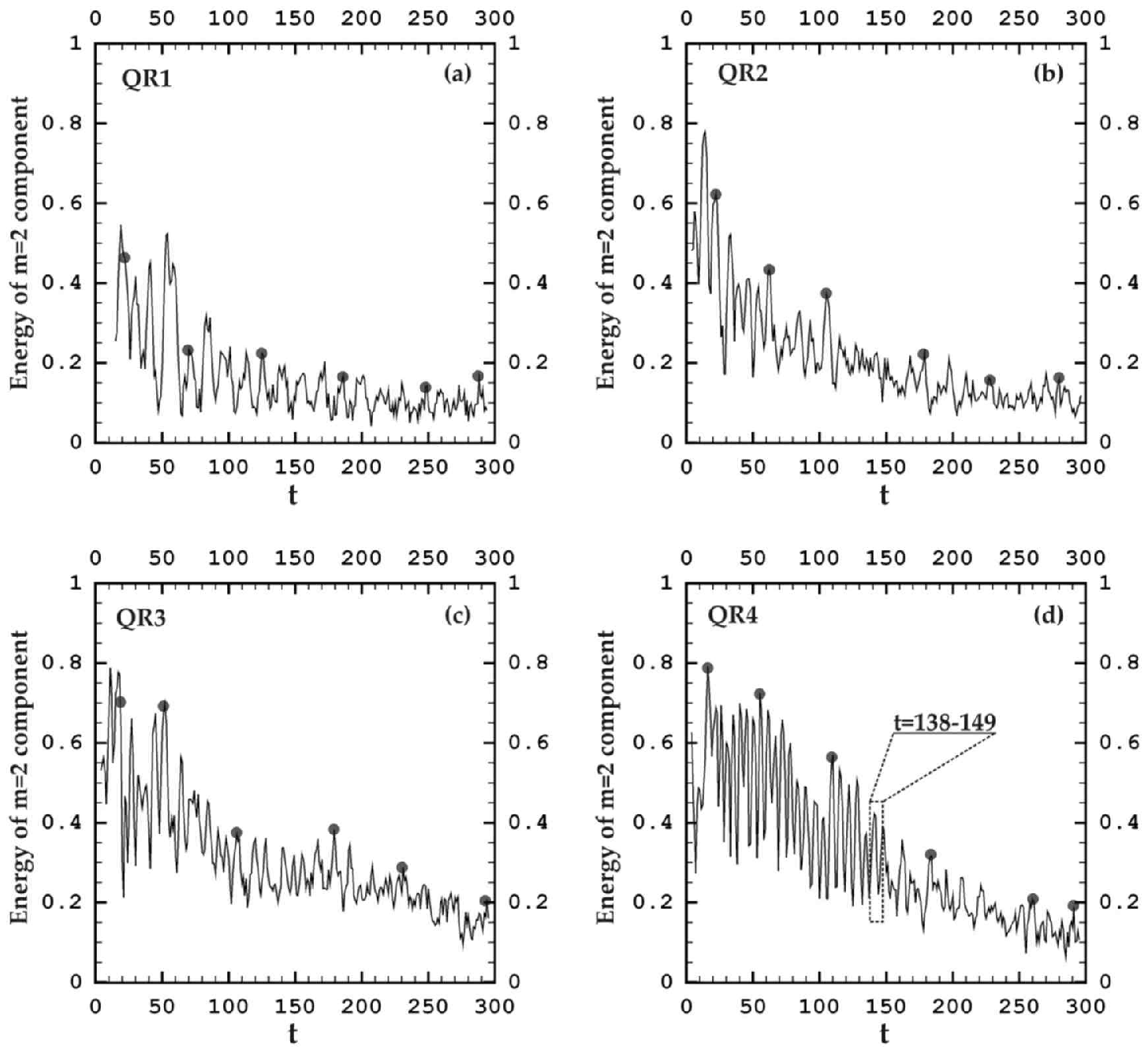}}
\caption{Evolution of the mean relative power of the $m=2$ mode
$<a_2^2>(t)$ evaluated in the region with radii between the
corotation radius $r_c$ and $r_c+1$. This quantity oscillates
quasi-periodically in a time scale depending on the rotation
period of the bar and decreases on the average due to the growth
of noise and the higher Fourier modes. The dots on the top of six
peak corresponds to the six snapshots shown in Fig. 5. In
\textbf{(d)}, the box with dashed line sides gives the time window
of the snapshots of Fig. 11.}\label{fig10}
\end{figure*}
\clearpage

\begin{figure*}
\centerline{\includegraphics[width=14cm]{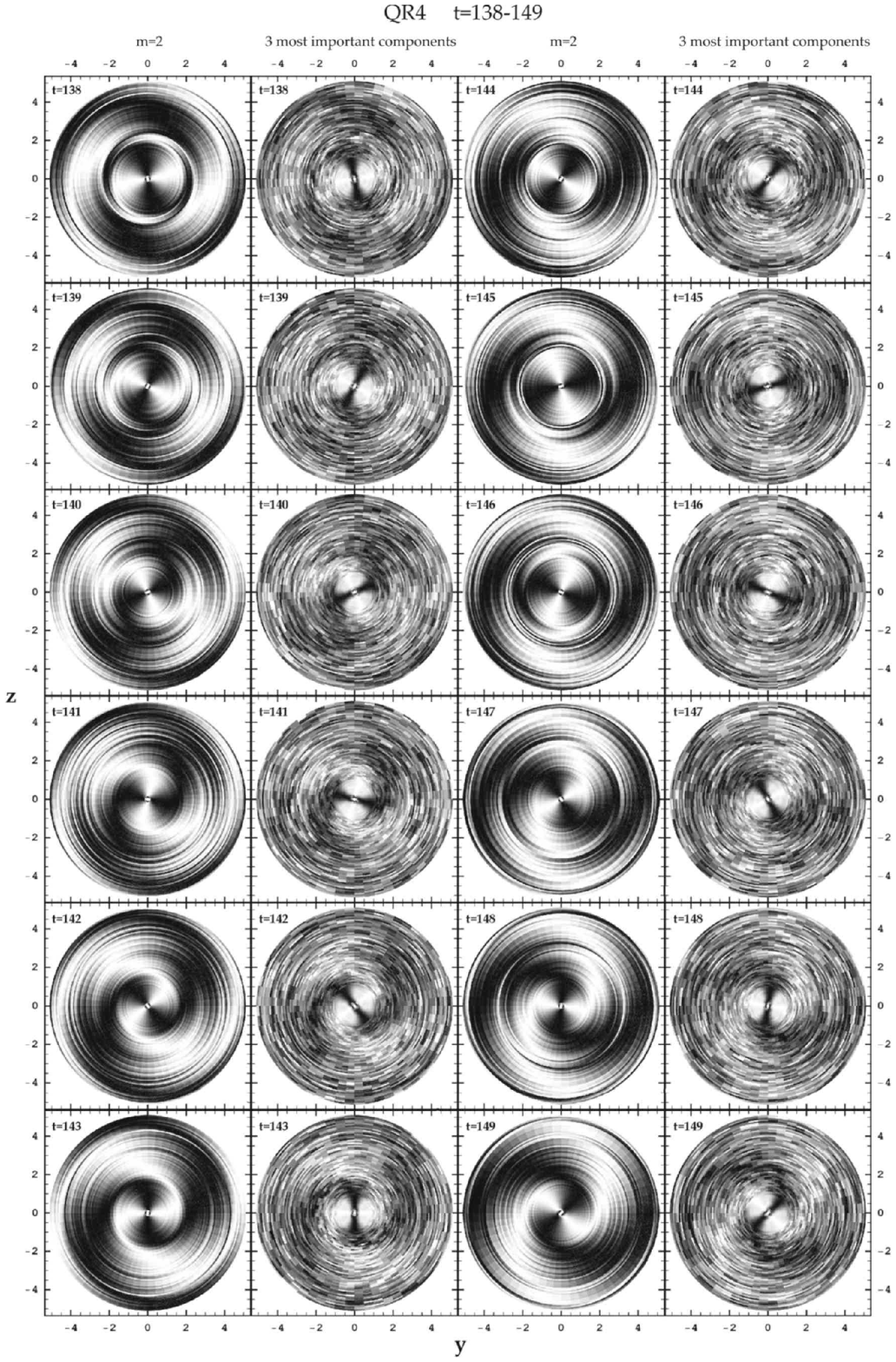}}
\caption{Detailed snapshots (every $\Delta t=1 T_{hmct}$) of
$\delta_2 (r, \phi, t)$ and $\delta_{3imp} (r, \phi, t)$ (in pairs
as in Fig. 5) for the model QR4, during the $18^{th}$-$19^{th}$
period of rotation of the bar. Observing carefully the evolution
of $\delta_2 (r, \phi, t)$, we see that a continuous spiral arm is
formed at $t=143$ (first column, last row) out to radii more than
$r=4$. At this time $<a_2^2>(t)$ has a maximum in Fig. 10d (inside
the box with dashed lines). At the next snapshots $t=144
\longrightarrow 147$ (third column) the bar goes temporarily out
of phase with respect to the previously excited wave, which
travels outwards and rotates more slowly than the bar. The
continuation of the spiral arms is temporarily broken. The bar
emits new waves travelling outwards that come in phase with the
previous wave at $t=149$ and form again a continuous spiral wave
extending out to large radii. This process is repeated up to the
end of the Hubble time, although spiral arms become fainter at
very late snapshots. The $m=2$ mode is discernible also in the
projections of $\delta_{3imp} (r, \phi, t)$ at the same snapshots
(second and fourth columns).}\label{fig11}
\end{figure*}
\clearpage

\begin{figure*}
\centerline{\includegraphics[width=\textwidth]{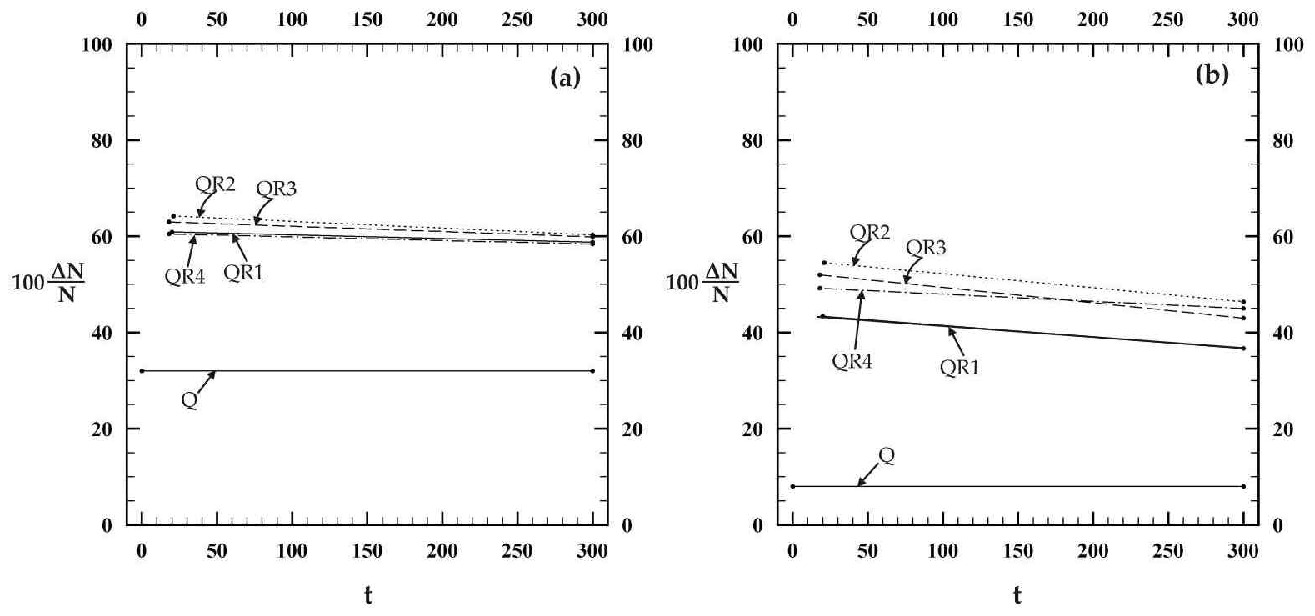}}
\caption{\textbf{(a)} The fraction of mass in chaotic motion
detected at $t_1\approx 20$ and at $t_2 =300$ for the all the
models. Rotation produces a serious increase of this fraction, (by
a factor of $\approx 2$, compare the non-rotating Q model with the
rotating models). \textbf{(b)} The fraction of mass in every
model, that can develop effective chaotic diffusion in a Hubble
time (with Lyapunov numbers $L_{cu} > 10^{-2}$). Rotation not only
increases the total chaotic mass, but also produces a shift in the
Lyapunov numbers, so that this fraction increases by a factor of 6
or 7 in the rotating models.}\label{fig12}
\end{figure*}
\clearpage

\begin{figure*}
\centerline{\includegraphics[width=\textwidth]{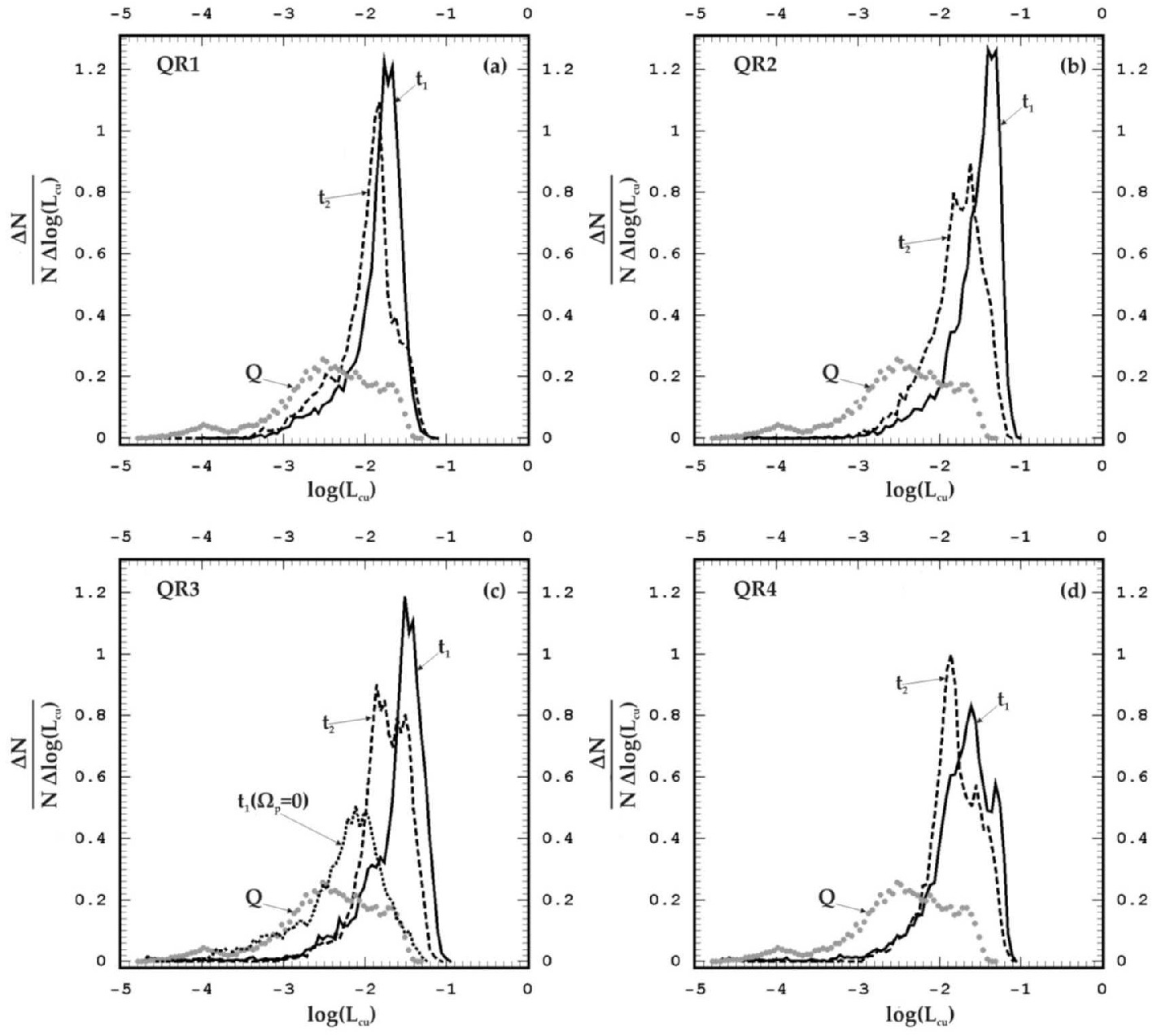}}
\caption{Distributions of the chaotic mass along the $\log L_{cu}$
axis normalized with respect to the total mass. The solid line
gives this distribution at $t_1$, while the dashed line at $t_2$
for every rotating model. The dotted line with index Q corresponds
to the non-rotating model. The transition from the curve Q to the
curve $t_1$ is due to the rotation. The dotted line with index
$t_1(\Omega_p=0)$ in \textbf{(c)} gives the chaotic mass in a
hypothetical model with the same self-consistent potential as QR3,
but not rotating ($\Omega_p=0$). The transition from the curve Q
to $t_1(\Omega_p=0)$ is only due to the different geometry of the
distribution of the mass caused by the rotation. The transition
from the curve $t_1(\Omega_p=0)$ to the curve $t_1$ is due to the
increase of the chaotic mass and to the shift of the Lyapunov
numbers due to the resonant effects. The transition from the curve
$t_1$ to the curve $t_2$ shows a slow tendency of the system for
self-organization.}\label{fig13}
\end{figure*}
\clearpage

\begin{figure}
\centerline{\includegraphics[width=8.5cm]{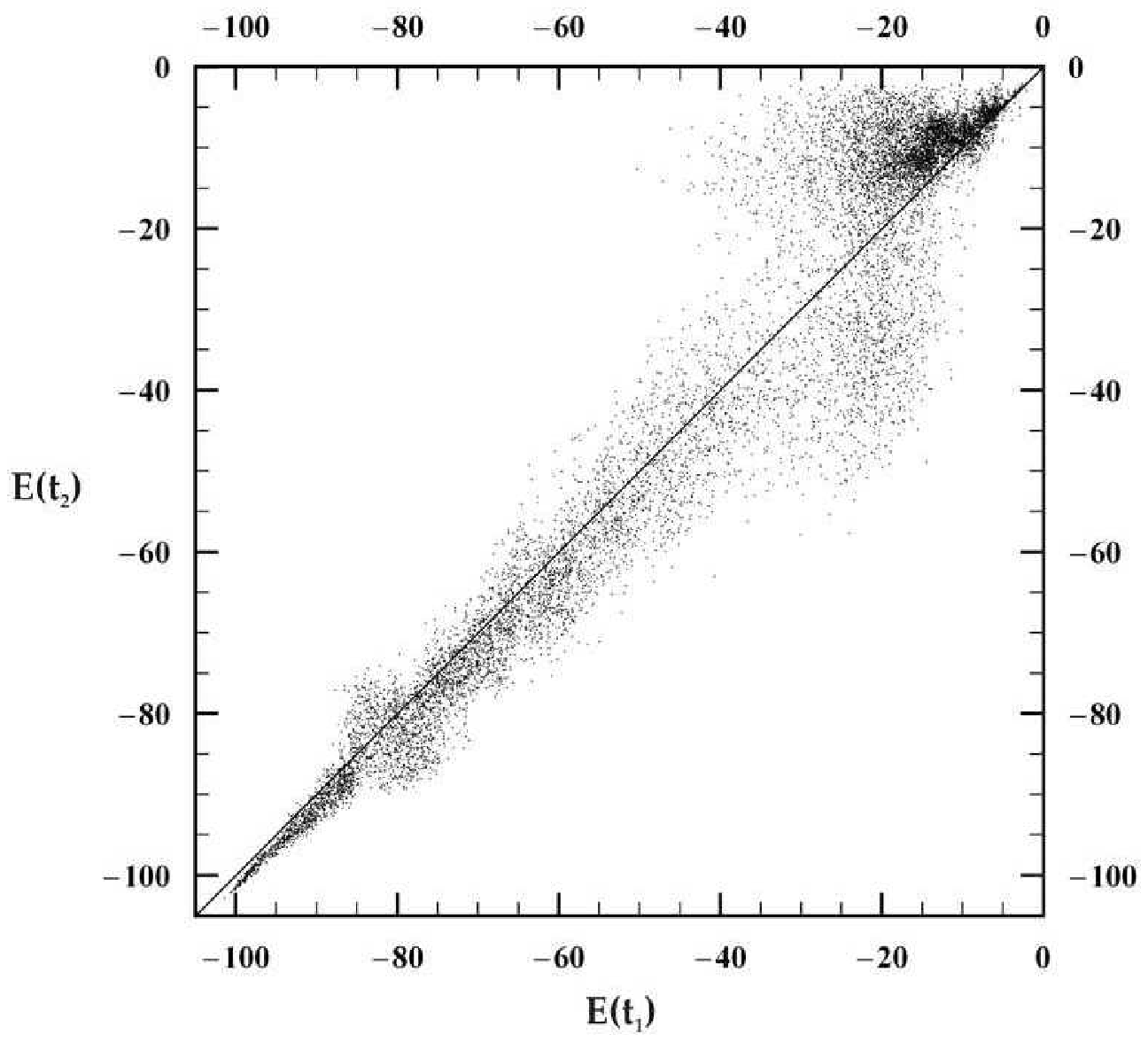}}
\caption{The binding energies of the particles $E(t_1)$ at $t_1$
vs the binding energies of the particles $E(t_2)$ at $t_2$ in QR4.
During the evolution from $t_1$ to $t_2$ particles at lower
energies lose energy, while particles at higher energies gain
energy. This allows self-organization in the inner parts, although
the total entropy of the system is expected to increase due to the
second law of thermodynamics.}\label{fig14}
\end{figure}
\clearpage

\begin{figure*}
\centerline{\includegraphics[width=14.5cm]{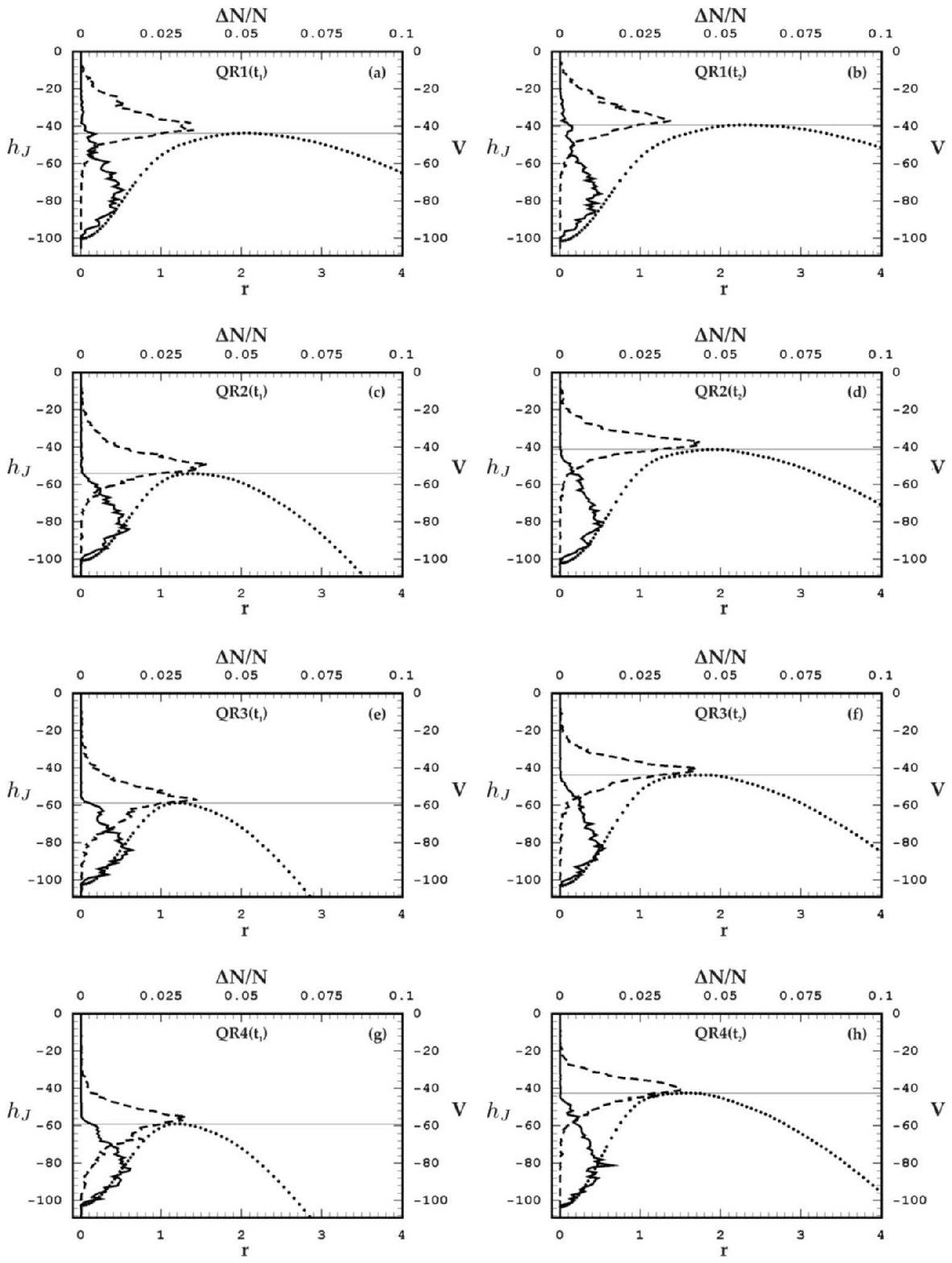}}
\caption{Distribution of the mass (upper horizontal axis) in
regular motion (solid line) and in chaotic motion (dashed line)
along the Jacobi integral $h_J$ (left vertical axis). The dotted
line gives the effective potential (right vertical axis) as  a
function of the radius $r$ (lower horizontal axis) taken along the
longest axis of the bar. The left column refers to the rotating
models at $t_1$ and the right column at $t_2$. The values of the
potential are normalized so that the deepest value at the center
of QR1 is $-100$. Regular orbits always dominate in the inner
parts of the system, well inside the corotation radius. Chaotic
orbits are always concentrated near corotation. The peak of their
distribution is always slightly above the Lagrangian points
$L1$,$L2$.} \label{fig15}
\end{figure*}
\clearpage

\begin{figure*}
\centerline{\includegraphics[width=14.5cm]{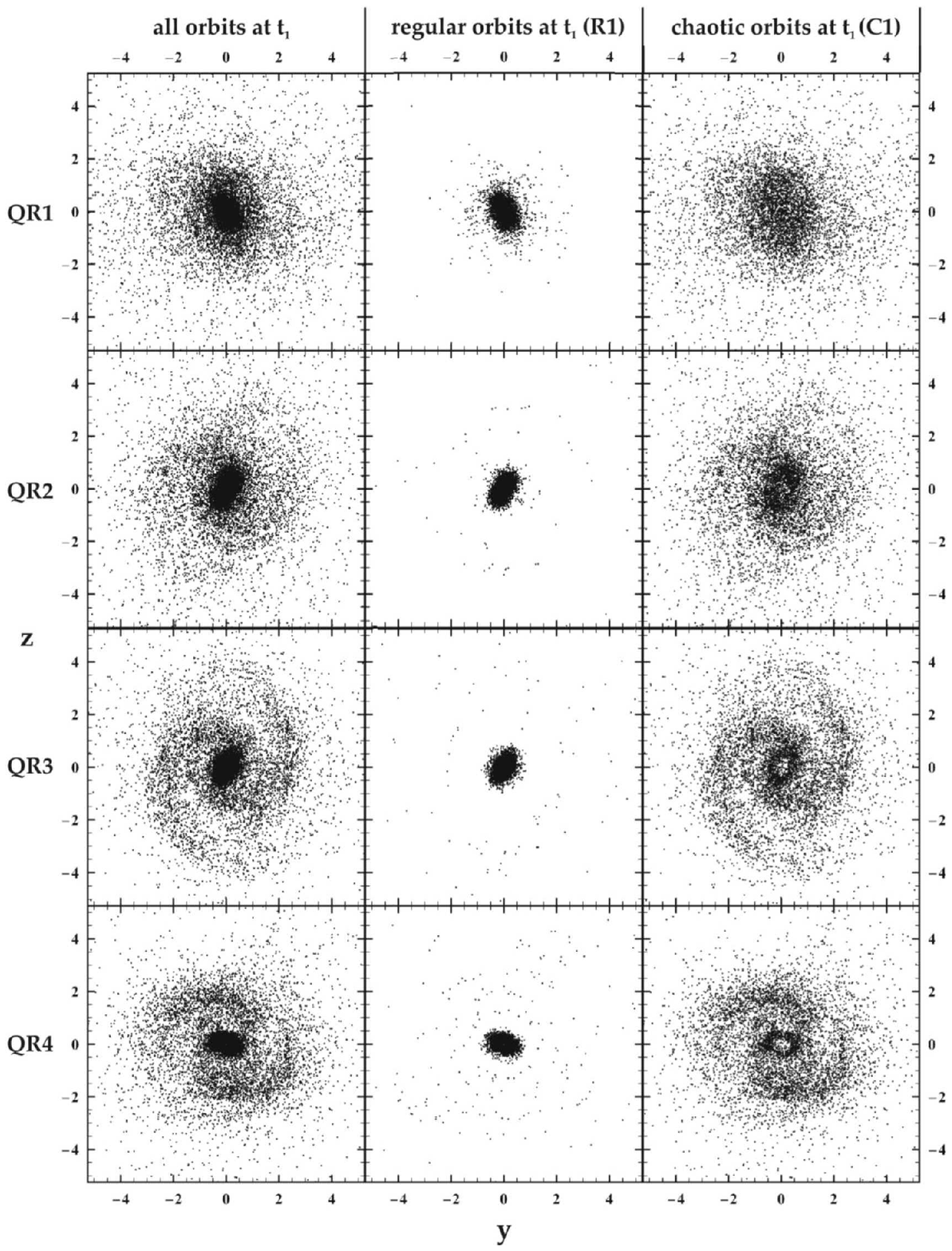}}
\caption{Projections on the equatorial plane of the four models at
$t_1 \approx 20$. Every row belongs to the same model. A sample of
1 every 12 particles uniformly distributed along the mass of the
model are plotted. The middle panel in every row shows the spatial
distribution of the particles in regular orbits (R1), while the
right panel shows the particles in chaotic orbits (C1) of this
sample. A spiral pattern (more clearly-sighted in QR2, QR3, QR4)
is almost completely formed by chaotic orbits.}\label{fig16}
\end{figure*}
\clearpage

\begin{figure*}
\centerline{\includegraphics[width=14.5cm]{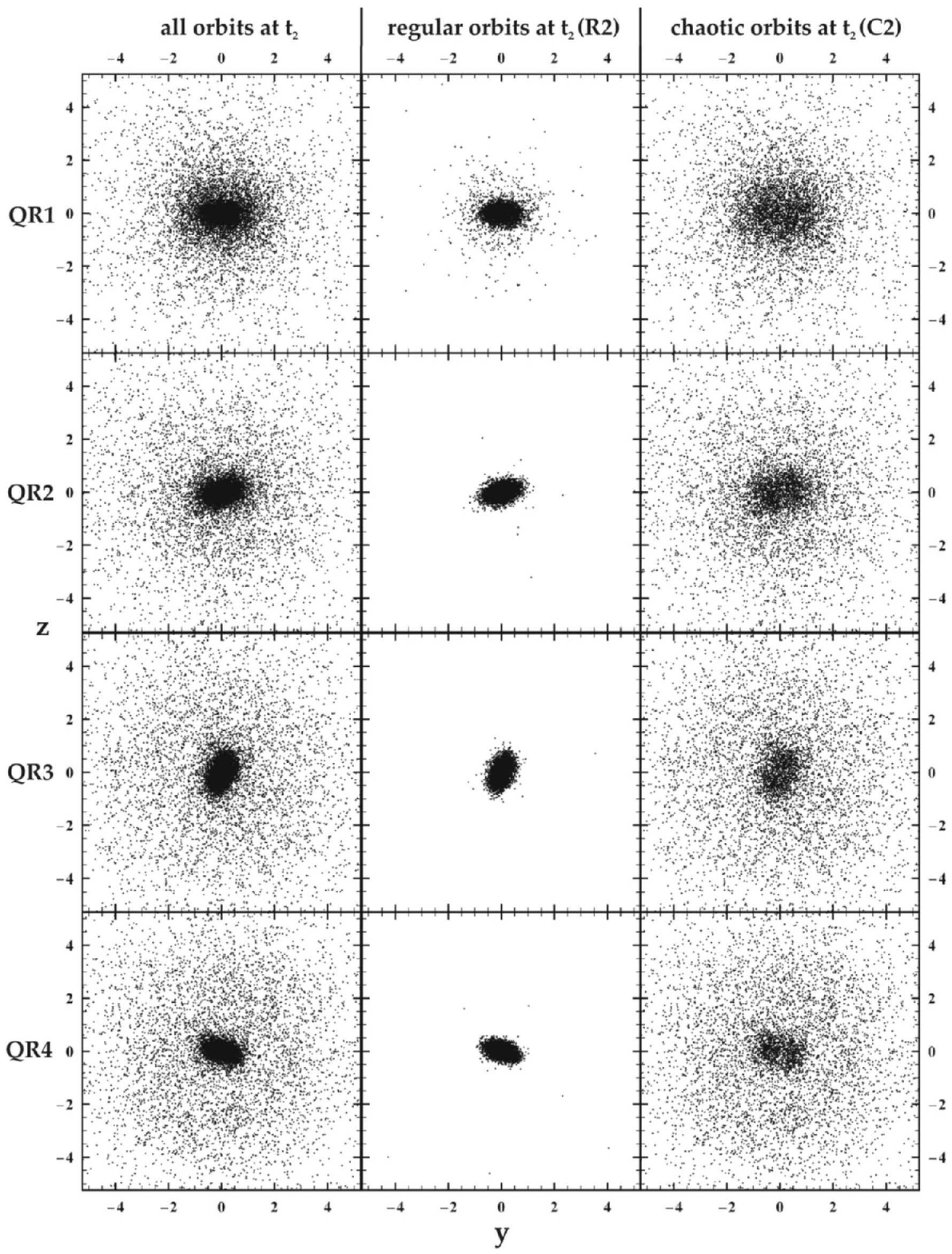}}
\caption{As in Fig. 16, but at $t_2=300$. The sets of particles in
regular and in chaotic motion found at this snapshot  are denoted
R2 and C2. About 91\%-93\% of the identities of the particles in
these sets maintain their initial character of motion and belong
also to the sets R1 and C1, respectively. Although a spiral
pattern cannot be easily seen in these projections, due to the
small amplitude of the $m=2$ mode, this mode preserves its spiral
pattern even at so late snapshots (see, for example, Fig. 8).
}\label{fig17}
\end{figure*}
\clearpage

\label{lastpage}

\begin{thebibliography}{99}


\bibitem{}Allen A.J., Palmer P.L., Papaloizou J., 1990, MNRAS, 242,576.

\bibitem{}
Athanassoula E. 2002, ApJ, 569, L83.

\bibitem{}Athanassoula E. 2003, MNRAS,  341, 1179.

\bibitem{}Barnes J., Efstathiou G., 1987, ApJ, 319, 575.

\bibitem{}Binney J., Tremaine S., 1987, Galactic Dynamics. Princeton
Series in Astrophysics.

\bibitem{} Block D.L., Buta R., Knapen J.H., Elmegreen D.M., Elmegreen B.G., Puerari
I., 2004, AJ, 128, 183

\bibitem{} Buta R., Block D.L., Knapen J.H., 2003, AJ, 126, 1148

\bibitem{} Buta R., Vasylyev S., Salo H., Laurikainen E., 2005, AJ, 130,
506

\bibitem{} Bournaud F., Combes F., 2002, A\&A, 392, 83

\bibitem{} Chirikov B.V., 1979, Phys. Rep., 52, 263

\bibitem{} Chirikov B.V., Israilev P.M., Shepelyansky D.L., 1971,
Physica D, 33, 77

\bibitem{} Contopoulos G., 1966, ApJ, 144, 1260

\bibitem{} Contopoulos G., 2004, Order and Chaos in Dynamical
Systems. Springer.

\bibitem{}Debattista V.P., Sellwood, J.A., 1998, ApJ, 493, L5.

\bibitem{}Debattista V.P., Sellwood, J.A., 2000, ApJ, 543, 704.

\bibitem{} Efstathiou G., Jones B.J.T., 1979, MNRAS, 186, 133.

\bibitem{} Fridman A.M., Khoruzhii O.V., Polyachenko, E.V.,
2002, Space Science Reviews, 102, 51

\bibitem{} Gadotti D.A., de Souza R.E., 2005, ApJ, 629, 797

\bibitem{} Gadotti D.A., de Souza R.E., 2006, ApJS, 163, 270

\bibitem{} Gnedin O.Y., Goodman J., Frei Z., 1995, AJ, 110, 1105

\bibitem{} Grosbol P.J., Patsis P., 1998, A\&A, 336, 840.

\bibitem{} Innanen K., 1966, ApJ, 143, 150.

\bibitem{} Kalapotharakos C., Voglis N., 2005, Cel. Mech. Dyn. Astr.,
92, 157.

\bibitem{} Kalapotharakos C., Voglis N., Contopoulos G., 2004, A\&A.,428, 905.

\bibitem{} Kormendy J., Kennicutt R.C. Jr., 2004, ARA\&A, 42, 603

\bibitem{} Lin C.C., Shu F.H., 1964, ApJ, 140, 646.

\bibitem{} Lin C.C., Shu F.H., 1966, Proc. Nat. Acad. Sci., 55,
229.

\bibitem{} Linden-Bell D., Kalnajs A.J., 1972, MNRAS, 157, 1.

\bibitem{} Lynden-Bell D., Wood R., 1968, MNRAS, 138, 495

\bibitem{} Muzzio J.C., Carpintero D.D., Wachlin F.C., 2005, Cel. Mech. Dyn.
Astr., 91, 173.

\bibitem{} Ohta K., Hamabe M., Wakamatsu K., 1990, ApJ, 357, 710.

\bibitem{} Peebles P.J.E., 1969, ApJ., 155, 393.

\bibitem{} Polyachenko E.V., 2002a, MNRAS, 330, 105

\bibitem{} Polyachenko E.V., 2002b, MNRAS, 331, 394

\bibitem{} Polyachenko V.L., Polyachenko E.V., 2002, Astronomy
Reports, 46, 1

\bibitem{} Rosenbluth M.N., Sagdeev R.A., Taylor J.B., Zaslavsky
G.M., 1966, Nucl. Fusion, 6, 217

\bibitem{} Shen J., Sellwood J. A., 2004, ApJ, 604, 614

\bibitem{} Skokos Ch., 2001, J. Phys. A: Math. Gen., 34, 10029.

\bibitem {} Tremaine S., Weinberg M., 1984, MNRAS, 209, 729.

\bibitem{} Voglis N., 1994, MNRAS, 267, 379

\bibitem{} Voglis N., 2003, MNRAS, 344, 575

\bibitem{} Voglis N., Hiotelis N., 1989, A\&A, 218,1.

\bibitem{} Voglis N., Hiotelis N., Hoflich P. 1991, A\&A, 249,5

\bibitem{} Voglis N., Contopoulos G., Efthymiopoulos C., 1998, Phys. Rev. E,
57, 372

\bibitem{} Voglis N., Contopoulos G., Efthymiopoulos C., 1999, Cel. Mech.
Dyn. Astr., 73, 211

\bibitem{} Voglis N., Kalapotharakos C., Stavropoulos I., 2002, MNRAS, 337, 619

\bibitem{} Voglis N., Tsoutsis P., Efthymiopoulos C., 2006,
MNRAS, submitted, (astro-ph/0607174)

\bibitem{} Walker G.H., Ford, J., 1969, Phys. Rev., 188, 416

\bibitem{} Weinberg M., 1985, MNRAS, 213, 451.

\bibitem{} Zaslavsky G.M., Chirikov B.V., 1972, Sov. Phys.
Upsekhi, 14, 549

\bibitem{} Zurek W.H., Quinn P.J., Salmon J.K., 1988, ApJ, 330, 519


\vspace{1cm}




\end{thebibliography}
\end{document}